\documentclass[10pt,a4paper]{iopart}
\usepackage{iopams}
\usepackage[english]{babel}
\usepackage{float}
\usepackage[dvips]{graphicx}
\usepackage{rotating}
\usepackage{epsfig}
\begin{document}
\title{Neutrino physics from precision cosmology}
\author{Steen Hannestad$^1$}
\ead{sth@phys.au.dk}
\address{$^1$Department of Physics and Astronomy, University
of Aarhus, Ny Munkegade, DK-8000 Aarhus C, Denmark }
\date{\today}

\begin{abstract}
Cosmology provides an excellent laboratory for testing various aspects of neutrino physics. Here, I review the current status of cosmological searches for neutrino mass, as well as other properties of neutrinos. Future cosmological probes of neutrino properties are also discussed in detail.
\end{abstract}
\maketitle

\section{Introduction}

The last few years has seen a dramatic increase in the amount and precision of cosmological data. Very precise measurements of the cosmic microwave background has been performed by the WMAP satellite \cite{kom10} and a number of different ground based experiments, most notably the ACBAR \cite{acbar} and QUAD \cite{quad} telescopes. The large scale distribution of galaxies has been measured by the Sloan Digital Sky Survey (SDSS) project which released its final data in 2009. A number of other experiments have probed the formation of cosmic structure using other techniques, some of which are still relatively new.
Altogether, the cosmological data is now precise enough that it is possible to probe various aspects of particle physics using cosmology. For example the density of dark matter has been measured quite precisely and since the dark matter particle is likely to be associated with physics at the TeV scale, i.e.\ with the energy scale probed at the LHC, cosmology provides an important second window to physics at this energy scale.

However, perhaps the best example of the interplay between cosmology and particle physics is in neutrino physics. Here, precision cosmology can be used to probe questions traditionally investigated in laboratory experiments.
While cosmology is at present not sensitive to the neutrino mass differences, it is highly sensitive to the absolute neutrino mass. Furthermore, this parameter is notoriously difficult to probe in laboratory experiments.
During the past 7-8 years neutrino cosmology has been a source of rapidly growing interest and a large number of papers have been written on the subject. There are also several good review articles available in the subject (e.g.\ \cite{Lesgourgues:2006nd,dolgov}) focusing on various aspects of neutrino cosmology. In this review I focus mainly on the effects of neutrinos in cosmic structure formation, with particular emphasis on the observational probes which can be used to measure neutrino masses.

Section 2 discusses the physics of neutrino decoupling in the early universe, including the effects of neutrinos on Big Bang nucleosynthesis. Section 3 briefly discusses the fact that during cosmic structure formation, long after neutrino decoupling, cosmic neutrinos are pure mass eigenstates and retain no flavour memory. In section 4 I review cosmic structure formation in some detail including both linear perturbation theory as well as neutrinos in non-linear structure formation. Section 5 is a review of the current constraints on neutrino mass, and section 6 contains a discussion of the current bound on the total energy density in neutrinos or other light, weakly interacting species.
Section 7 discusses various future observational probes as well as their potential and possible systematics, and section 8 contains a brief discussion of the cosmic neutrino background anisotropy.
Finally, section 9 contains a brief discussion and conclusion.

\subsection{Neutrino mixing}

Neutrino oscillation experiments have shown beyond any reasonable doubt that neutrinos have non-zero masses. Furthermore, two distinct mass differences have been measured in different types of experiments, perfectly compatible with the assumption that there are three distinct mass eigenstates, $\nu_M$, corresponding two the three known flavour eigenstates, $\nu_F$. These states are connected via
\begin{equation}
\nu_F = U \nu_M,
\end{equation}
where $U$ is a $3 \times 3$ unitary matrix. For Dirac neutrinos the mixing matrix contains 4 free parameters, the three rotation angles, and one complex phase, $\delta$. For Majorana neutrinos there are two additional phases which cannot be rotated away. These Majorana phases have no influence on oscillation observables, but are important for example in neutrinoless double beta decay (see e.g.\ \cite{Fukugita:2003en,Giunti:2007ry} for details).

The matrix $U$ can be parameterized in a number of ways, but the most popular by far is in terms of the mixing angles and the phases. In the relativistic limit neutrino oscillation probabilities only depend on the masses of the three states via $\Delta m^2$, i.e.\ in addition to the elements of the mixing matrix oscillation experiments are sensitive to two $\Delta m^2$ values.

By now constraints on $U$ and $\Delta m^2$ come from a variety of different experiments, the most recent overview is given in Ref.~\cite{GonzalezGarcia:2010er}
and provides the following constraints (at 3$\sigma$)
\begin{eqnarray}
\theta_{12} & = & (34 \pm 3)^\circ \\
\theta_{23} & = & (43^{+11}_{-7})^\circ \\
\theta_{13} & = & \leq 12.5^\circ \\
\delta m^2_{12} & = & 7.6 \pm 0.6 \times 10^{-5} \, {\rm eV}^2 \\
\delta m^2_{31} & = & \pm (2.4 \pm 0.4) \times 10^{-3} \, {\rm eV}^2 \\
\delta_{\rm CP} & = & {\rm No \,\, constraint}
\end{eqnarray}

As will be discussed later, apart from light element abundances, cosmological observables in most cases depend little on flavour and conversely cosmology provides little information on the mixing structure of neutrinos.

\subsection{Neutrino masses}

While oscillation experiments are by far the most useful probe of the mixing angles and the mass {\it differences}, $\Delta m^2$, they are hardly useful for probing the absolute neutrino masses, $m_i$.
To measure these parameters there are currently three useful paths. One possibility is to look for neutrinoless double beta decay which can occur because the emitted left handed neutrino contains a negative helicity component with relative amplitude $m/E$, which can then be absorbed as an antineutrino, provided that neutrinos are Majorana particles. The amplitude for this process is proportional to $|\sum_j U_{ej}^2 m_j|$ (see e.g.\ \cite{Giunti:2007ry} for for details). This particular form comes from the fact that each neutrino vertex is proportional to $U_{ej}$ and only the ``opposite'' helicity component contributes. Note that the sum retains the phase structure so that phase cancelation is in principle possible.

The actual parameter measured in any neutrinoless double beta decay
experiment is the half-life $T_{1/2}$ for some isotope which is related to $m_{\beta
\beta} = |\sum_j U_{ej}^2 m_j|$ via the relation
\begin{equation}
\frac{1}{T_{1/2}} = G^{0\nu} \left|M^{0\nu}\right|^2 m_{\beta
\beta}^2,
\end{equation}
where $G^{0\nu}$ is a phase-space factor and
$\left|M^{0\nu}\right|^2$ the nuclear matrix element squared. Translating any measurement of a finite lifetime
to an effective neutrino mass therefore involves an uncertainty related to the calculation of the nuclear
matrix element (see e.g.\ \cite{Rodin:2006yk,Rodin:2007fz} for a discussion of this point). The best upper bound on $m_{\beta \beta}$ from double beta decay currently comes from the Heidelberg-Moscow (HM) experiment and is
is $m_{\beta \beta} < 0.27 \, {\rm eV}$ (90\% C.L.) \cite{KlapdorKleingrothaus:2000sn,Rodin:2007fz}.

However, there is also a claim of a positive signal for the decay in a different analysis of the (HM) experiment
\cite{Klapdor-Kleingrothaus:2001ke,Klapdor-Kleingrothaus:2004wj,%
VolkerKlapdor-Kleingrothaus:2005qv}.
The upcoming GERDA \cite{Schonert:2005zn} and EXO \cite{Akimov:2005mq} experiments which are currently in commissioning will have the capability to either confirm or rule out this claim.

Another option is to look for the kinematical effect of a non-zero neutrino mass in ordinary beta decay. This is challenging, not because the process is forbidden for zero neutrino mass, but because the neutrino mass only produces a noticeable effect very close to the endpoint of the electron spectrum where there are very few events.
In the beta decay vertex, an electron neutrino is emitted. However, since the subsequent measurement involves the electron energy it is more convenient to view the beta decay as three separate possible processes which contain either $\nu_1,\nu_2$ or $\nu_3$. The beta decay rate is the incoherent sum of these three processes, and the rate therefore involves $\sum U_{ej}^* U_{ej}$, i.e.\ it does not retain any Majorana phase information.
The shape factor for the electron decay in the presence of more than one massive neutrino is given by
\begin{equation}
S(E_e) \propto (Q-E_e)\sum_j U_{ej}^* U_{ej} \sqrt{(Q-E_e)^2-m_j^2}.
\end{equation}
As long as the energy resolution of the experiment is significantly worse than the mass splittings this can be written as
\begin{equation}
S(E_e) \propto (Q-E_e)\sqrt{(Q-E_e)^2 - \sum_j U_{ej}^* U_{ej} m_j^2},
\end{equation}
i.e.\ the spectrum distortion can be described by a single effective mass. The current upper bound on this effective
mass is
\begin{equation}
m_\beta = \left(\sum_j U_{ej}^* U_{ej} m_j^2\right)^{1/2} < 2.3 \, {\rm eV}
\end{equation}
at 95\% C.L. from the final analysis of the Mainz experiment \cite{kraus}. Starting in early 2012 the KATRIN experiment \cite{guido,katrin} will improve this sensitivity by an order of magnitude to 0.2 eV.

The final method is to look for the kinematic effect of neutrino masses in cosmological structure formation. As will be discussed in detail in the next sections, a non-zero neutrino mass means that neutrinos contribute to the present density of dark matter. However, since their thermal history is very different from that of cold dark matter they have a very distinct signature on cosmic structure.
As a first approximation, cosmology is sensitive to the total energy density in neutrinos, which for non-relativistic neutrinos is simply proportional to the sum of all neutrino mass eigenstates, $\sum m_j$. However, this is true if the sensitivity of cosmological data to the neutrino mass is poor compared with the internal mass splittings, i.e.\ if $\sigma(m) \gtrsim \sum m_\nu$. Some proposed cosmological structure surveys in the coming decade will be sensitive to masses as low as 0.03-0.05 eV, which means that the effect of massive neutrinos cannot simply be approximated as a sum of the involved mass eigenstates. This possibility will be discussed later.

The complementarity of these three ways to probe neutrino masses has been discussed in some detail in the literature. While they all probe the neutrino masses, they do so via different combinations of the mass states with the mixing matrix elements.
They are also prone to completely different systematics, and, just as important, to different modeling uncertainties. For example, the presence of right-handed currents will affect both the neutrinoless double beta decay rate and the shape of the beta decay spectrum, but have no impact on the cosmological mass measurements. It is therefore entirely possible that an experiment like KATRIN measures a spectrum distortion, while cosmology provides no signature.

\section{Neutrino decoupling}

Before going on to discuss the impact of neutrinos on structure formation I will briefly review the standard picture of neutrino decoupling in the early universe.

In the standard model neutrinos interact via weak interactions
with charged leptons, keeping them in equilibrium with the
electromagnetic plasma at high temperatures. Below $T \sim 30-40$
MeV $e^+$ and $e^-$ are the only relevant particles, greatly
reducing the number of possible reactions which must be
considered. In the absence of oscillations neutrino decoupling can
be followed via the Boltzmann equation for the single particle
distribution function \cite{kolb}
\begin{equation}
\frac{\partial f}{\partial t} - H p \frac{\partial f}{\partial p}
= C_{\rm coll}, \label{eq:boltz}
\end{equation}
where $C_{\rm coll}$ represents all elastic and inelastic
interactions. In the standard model all these interactions are $2
\leftrightarrow 2$ interactions in which case the collision
integral for process $i$ can be written
\begin{eqnarray}
C_{\rm coll,i} (f_1) & = & \frac{1}{2E_1} \int \frac{d^3 {\bf
p}_2}{2E_2 (2\pi)^3} \frac{d^3 {\bf p}_3}{2E_3 (2\pi)^3} \frac{d^3
{\bf p}_4}{2E_4 (2\pi)^3} \nonumber \\
&& \,\, \times (2\pi)^4 \delta^4
(p_1+p_2-p_3+p_4)\Lambda(f_1,f_2,f_3,f_4) S |M|^2_{12 \to 34,i},
\end{eqnarray}
where $S |M|^2_{12 \to 34,i}$ is the spin-summed and averaged
matrix element including the symmetry factor $S=1/2$ if there are
identical particles in initial or final states. The phase-space
factor is $\Lambda(f_1,f_2,f_3,f_4) = f_3 f_4 (1-f_1)(1-f_2) - f_1
f_2 (1-f_3)(1-f_4)$.

The matrix elements for all relevant processes can for instance be
found in Ref.~\cite{Hannestad:1995rs} (however, see also \cite{dolgov}). If Maxwell-Boltzmann
statistics is used for all particles, and neutrinos are assumed to
be in complete scattering equilbrium so that they can be
represented by a single temperature, then the collision integral
can be integrated to yield the average annihilation rate for a
neutrino
\begin{equation}
\Gamma = \frac{16 G_F^2}{\pi^3} (g_L^2 + g_R^2) T^5,
\end{equation}
where
\begin{equation}
g_L^2 + g_R^2 = \cases{\sin^4 \theta_W + (\frac{1}{2}+\sin^2
\theta_W)^2 & for $\nu_e$ \cr \sin^4 \theta_W +
(-\frac{1}{2}+\sin^2 \theta_W)^2 & for $\nu_{\mu,\tau}$}.
\end{equation}

This rate can then be compared with the Hubble expansion rate
\begin{equation}
H = 1.66 g_*^{1/2} \frac{T^2}{M_{\rm Pl}} \label{eq:hub}
\end{equation}
to
find the decoupling temperature from the criterion $\left. H =
\Gamma \right|_{T=T_D}$. From this one finds that $T_D(\nu_e)
\simeq 2.4$ MeV, $T_D(\nu_{\mu,\tau}) \simeq 3.7$ MeV, when $g_*
=10.75$, as is the case in the standard model.

This apparently straightforward conclusion is complicated by neutrino oscillations. As has been shown in a number of papers the mass differences and mixing angles of the neutrino sector are such that all species are effectively almost equilibrated prior to neutrino decoupling. To a good approximation neutrinos can therefore be treated as a single species with an averaged coupling strength, decoupling slightly prior to the estimated decoupling temperature of electron neutrinos.

A coupling temperature of around 2.5-3 MeV
means that neutrinos decouple at a temperature which is
significantly higher than the electron mass. When $e^+e^-$
annihilation occurs around $T \sim m_e/3$, the neutrino
temperature is unaffected whereas the photon temperature is heated
by a factor $(11/4)^{1/3}$. The relation $T_\nu/T_\gamma =
(4/11)^{1/3} \simeq 0.71$ holds to a precision of roughly one
percent. The main correction comes from a slight heating of
neutrinos by $e^+e^-$ annihilation, as well as finite temperature
QED effects on the photon propagator
\cite{Dicus:1982bz,Rana:1991xk,herrera,Dolgov:1992qg,Dodelson:1992km,%
Fields:1993zb,Hannestad:1995rs,Dolgov:1997mb,Dolgov:1999sf,gnedin,%
Esposito:2000hi,Steigman:2001px,Mangano:2001iu,osc,Mangano:2005cc}.

The extra energy deposited in neutrinos is usually defined in terms of an extra number of neutrino species, $\Delta N_\nu$, defined as
\begin{equation}
\Delta N_\nu = \frac{\Delta \rho}{\rho_{\nu,0}},
\end{equation}
where $\rho_{\nu,0}$ is the energy density in a single neutrino species assuming complete decoupling.

The most precise calculation to date \cite{Mangano:2005cc} has estimated that $\Delta N_\nu = 0.046$, so the standard model prediction for the total energy density in neutrinos is $N_\nu = 3.046$. As will be discussed later the difference of 0.046 is probably too small to be detectable even with future observational data.

It should be noted here that $N_\nu$ is customarily used in cosmology to parameterize {\it any} additional relativistic energy density, not just neutrinos. Therefore $N_\nu$ is one of the cosmological parameters normally fitted in cosmological parameter estimation and any value significantly different from 3 could indicate the presence of new physics beyond the standard model.

\subsection{Big Bang nucleosynthesis and the number of neutrino species}

Shortly after neutrino decoupling the weak interactions which keep
neutrons and protons in statistical equilibrium freeze out. Again
the criterion $\left. H = \Gamma \right|_{T=T_{\rm freeze}}$ can
be applied to find that $T_{\rm freeze} \simeq 0.5 g_*^{1/6}$ MeV
\cite{kolb}.

Eventually, at a temperature of roughly 0.2 MeV deuterium starts
to form, and very quickly all free neutrons are processed into
$^4$He. The final helium abundance is therefore roughly given by
\begin{equation}
Y_P \simeq \left. \frac{2 n_n/n_p}{1+n_n/n_p} \right|_{T\simeq 0.2
\,\, {\rm MeV}}.
\end{equation}

$n_n/n_p$ is determined by its value at freeze out, roughly by the
condition that $n_n/n_p|_{T=T_{\rm freeze}} \sim
e^{-(m_n-m_p)/T_{\rm freeze}}$.

Since the freeze-out temperature is determined by $g_*$ this in
turn means that $g_*$ can be inferred from a measurement of the
helium abundance. However, since $Y_P$ is a function of both
$\Omega_b h^2$ and $g_*$ it is necessary to use other measurements
to constrain $\Omega_b h^2$ in order to find a bound on $g_*$.
Historically, this has been done by using the measured abundance of
deuterium as a probe of the cosmic baryon density. However, at present
the most precise determination of the baryon density by far comes
from the cosmic microwave background observations.

When the baryon density is fixed, $Y_P$ can be used to constrain $g_*$.
As will be seen later, observations of the cosmic microwave background
also provide a constraint on $g_*$ which is quite stringent and the
two types of observations can be used as a consistency check for the
standard radiation dominated expansion in the region ${\rm eV} < T < {\rm MeV}$.
Any discrepancy between the two values of $g_*$ could in principle indicate
non-standard physics such as the late decay of a massive particle.

Usually such bounds are expressed in
terms of the equivalent number of neutrino species, $N_\nu \equiv
\rho/\rho_{\nu_0}$, instead of $g_*$. The exact value of the BBN bound on $N_\nu$ is
somewhat uncertain because of systematic uncertainties involved in the $Y_P$
determination.

Very recent analyses are those found in \cite{Izotov:2010ca,Aver:2010wq} which indicate
$Y_P = 0.2565 \pm 0.001 \, ({\rm stat}) \pm 0.005 \, ({\rm syst})$ \cite{Izotov:2010ca}
and $Y_P = 0.2561 \pm 0.011$ \cite{Aver:2010wq} respectively (see also
\cite{Steigman:2010pa}). The central values are consistent
with $N_\nu \sim 3.7$, i.e.\ a value somewhat higher than the predicted
$N_\nu = 3.04$. Even though this might point to non-standard physics,
the possible systematics seem too large for any firm conclusions.
However, it is intriguing that CMB observations currently point in the same
direction, i.e.\ $N_\nu > 3$.

Another interesting parameter which can be constrained by the same
argument is the neutrino chemical potential, $\xi_\nu=\mu_\nu/T$
\cite{Kang:xa,Kohri:1996ke,Orito:2002hf,Ichikawa:2002vn}. At first
sight this looks like it is completely equivalent to constraining
$N_\nu$. However, this is not true because a chemical potential
for electron neutrinos directly influences the $n-p$ conversion
rate. Furthermore, it is crucial to take neutrino flavour
oscillations into account when calculating bounds on the neutrino
chemical potential. This point is discussed in more detail below.

\subsection{The effect of oscillations}

In the previous section the one-particle distribution function,
$f$, was used to describe neutrino evolution. However, for
neutrinos the mass eigenstates are not equivalent to the flavour
eigenstates because neutrinos are mixed. Therefore the evolution
of the neutrino ensemble is not in general described by the three
scalar functions, $f_i$, but rather by the evolution of the
neutrino density matrix, $\rho \equiv \psi \psi^\dagger$, the
diagonal elements of which correspond to $f_i$.

Using the density matrix, $\rho$, the Boltzmann equation is replaced
by (see e.g.\ \cite{McKellar:1992ja})
\begin{equation}
\frac{\partial \rho}{\partial t} - p H \frac{\partial \rho}{\partial p} =
-\left[\frac{{\cal M}}{2 p} - \frac{8 \sqrt{2} G_F p}{3 m_W^2},\rho \right] + C[\rho].
\label{eq:density}
\end{equation}
The first term accounts for oscillations and the last term is the ${\cal O} (G_F^2)$
collision operator. $\cal M$ is the mass matrix in flavour space, i.e.\
${\cal M} = U^\dagger M U$, with $M = {\rm diag}(m_1^2,m_2^2,m_3^2)$.

Eq.~\ref{eq:density} was solved approximately for the case of standard neutrinos
in Ref.~\cite{osc}, and exactly in \cite{Mangano:2005cc}.

Without oscillations it is possible to compensate a very large
chemical potential for muon and/or tau neutrinos with a small,
negative electron neutrino chemical potential \cite{Kang:xa}.
However, since neutrinos are almost maximally mixed a chemical
potential in one flavour can be shared with other flavours, and
the end result is that during BBN all three flavours have almost
equal chemical potential
\cite{lunardini,Pastor:2001iu,Dolgov:2002ab,Abazajian:2002qx,Wong:2002fa}.
This in turn means that the bound on $\nu_e$ applies to all
species. The most recent bound on the neutrino asymmetry from BBN
comes from \cite{Serpico:2005bc}

\begin{equation}
-0.04 \leq \xi_i = \frac{|\eta_i|}{T} \leq 0.07
\end{equation}
for $i=e,\mu,\tau$ \footnote{It should be noted that this bound is strictly speaking only valid assuming that there
is no extra relativistic energy density in other species present.}.

The bound assumes complete flavour equilibration during BBN, which
with the measured mixing angles and mass differences is a fairly
good approximation.
It should, however, be noted that with some fine tuning of the initial
conditions it is possible to have large lepton asymmetries while still
producing the correct light element abundances \cite{Pastor:2008ti}.
Another possibility is that additional majoron type interactions may
allow for large asymmetries to be present \cite{dolgov2004}.

In models where sterile neutrinos are present
even more remarkable oscillation phenomena can occur. However, I
do not discuss this possibility further, and instead
refer to the review \cite{dolgov}.

\section{What is a cosmic background neutrino?}

An interesting question which is not often addressed in the literature is whether cosmic background neutrinos are mass or flavour eigenstates, or neither.
Even if neutrinos are created as flavour states in charged current interactions, they stop interacting at the MeV scale and after this point they propagate freely.
The neutrino created in a charged current interaction can be thought of as a single wavepacket consisting of a superposition of three mass states. These three wavepackets travel at slightly different velocity and at some point the original wavepackets will no longer overlap, i.e.\ the original flavour state has decohered and separated into distinct wavepackets related to mass states.

The timescale for this decoherence phenomenon to occur can be estimated from the following argument
(see \cite{Giunti:2007ry,Farzan:2008eg} for a more detailed discussion).

At creation, the neutrino wave packet should have a size of roughly $\delta_x = c \Delta t_{\rm coll}$, where $\Delta t_{\rm coll}$ is the mean time between collisions for particles in the medium.
To be conservative we assume the particles in the medium to interact only via the weak interactions which create the neutrinos. In practise electrons for example also have electromagnetic interactions which are much faster. Assuming only weak interactions is therefore an upper bound on the time between collisions and therefore also an upper bound on the size of the emitted neutrino wavepackets.

At decoupling, by definition, the timescale for collisions is comparable to the Hubble time, $\Delta t_{\rm coll} \sim H^{-1} \sim m_{\rm Pl}/T_D^2$. At some later time the size of the wavepacket will have increased to $\delta_x = m_{\rm Pl}/(T_D T)$ because of cosmic expansion. Now assume that the original wavepacket is a superposition of two wavepackets related to different masses, $m_1$ and $m_2$. The velocity difference between the two packets is $\Delta v \sim (m_2^2-m_1^2)/p^2 = \delta m^2/p^2 \sim \delta m^2/T^2$, and once the neutrino has traveled a distance of $L \sim \delta_x/\Delta v$ the two wavepackets no longer overlap and the original state has decohered.
The distance covered by the neutrino at some given time is the Hubble scale at that time, $m_{\rm Pl}/T^2$, again assuming that neutrinos are ultrarelativistic. This distance is longer than the distance needed for decoherence provided that the decoherence condition
\begin{equation}
\delta m^2 > T^3/T_D ,
\end{equation}
is fulfilled. Note that this assumes radiation domination, in a matter dominated universe the condition is slightly different.
At matter-radiation equality this corresponds to $\delta m^2 > 10^{-6} \, {\rm eV}^2$ which is fulfilled by both $\delta m_{21}^2$ and $\delta m_{31}^2$. We again note that this is a very conservative estimate, in practise charged leptons have much shorter collision timescales, and $\delta p$ is therefore correspondingly much larger than the pure weak interaction estimate.

Thus, cosmic background neutrinos can be treated as exact mass eigenstates during the entire history of structure formation in the universe.

This conclusion also leads to the question: What is the number density of neutrino states of mass state $i$ in the universe? At high temperature long before neutrino decoupling, the neutrino density matrix is diagonal in flavour space, with the diagonal elements characterised by temperature and chemical potential, $T$ and $\xi_\alpha$.
After decoupling the evolution becomes simple in mass basis, i.e.\
\begin{equation}
U \dot \rho U^\dagger = \dot\rho_M = 0
\label{eq:mass}
\end{equation}
in the absence of decoherence. Decoherence can be approximated by a damping term in Eq.~\ref{eq:mass}, such that
\begin{equation}
\dot \rho_M = -K (\rho_M - {\rm diag}(\rho_M)),
\end{equation}
where $K$ is a temperature dependent decoherence rate.
In the absence of flavour asymmetry the evolution is unimportant, but in the presence of an asymmetry the final density of states 1,2 and 3 is given by
\begin{eqnarray}
n_i = \rho_{M,ii}(t=t_{\rm dec})
\end{eqnarray}
with all off-diagonal components in the mass basis damped away.

\section{Neutrinos in cosmic structure formation}

Neutrinos are a source of dark matter in the present day universe
simply because they contribute to $\Omega_m$. The present
temperature of massless standard model neutrinos is $T_{\nu,0} =
1.95 \, K = 1.7 \times 10^{-4}$ eV, and any neutrino with $m \gg
T_{\nu,0}$ behaves like a standard non-relativistic dark matter
particle.

The present contribution to the matter density of $N_\nu$ neutrino
species with standard weak interactions is given by
\begin{equation}
\Omega_\nu h^2 = N_\nu \frac{m_\nu}{94.57 \, {\rm eV}}
\end{equation}
Just from demanding that $\Omega_\nu \leq 1$ one finds the bound
\cite{Gershtein:gg,Cowsik:gh}
\begin{equation}
m_\nu \lesssim \frac{46 \, {\rm eV}}{N_\nu} \label{eq:mnu}
\end{equation}
More realistically, one could assume that all dark matter is in the form of neutrinos. Very loosely, the total
matter density is $\Omega_m \sim 0.3$, with approximately 0.05 in the form of baryons. A more realistic upper
bound would then be $\Omega_\nu h^2 \lesssim 0.12$, leading to
\begin{equation}
m_\nu \lesssim \frac{11 \, {\rm eV}}{N_\nu},
\end{equation}
a number comparable to the present upper limit from beta decay experiments. However, as will be seen below, observations of cosmic
structure allow for a much tighter constraint on the neutrino mass.

\subsection{The Tremaine-Gunn bound}

If neutrinos are the main source of dark matter, then they must
also make up most of the galactic dark matter. However, neutrinos
can only cluster in galaxies via energy loss due to gravitational
relaxation since they do not suffer inelastic collisions. In
distribution function language this corresponds to phase mixing of
the distribution function \cite{Tremaine:we}.
Initially, the microscopic neutrino distribution function is given by the simple non-degenerate Fermi-Dirac distribution
\begin{equation}
f(p) = \frac{1}{e^{p/T}+1},
\end{equation}
where $T = T_0/a$, i.e.\ the distribution function is preserved as an ultra-relativistic distribution function after neutrino decoupling in the early universe.

By using the theorem
that the phase-mixed or coarse grained distribution function must
explicitly take values smaller than the maximum of the original
distribution function one arrives at the condition
\begin{equation}
f_{\rm CG} \leq f_{\nu,{\rm max}} = \frac{1}{2}
\end{equation}
Because of this upper bound it is impossible to squeeze neutrino
dark matter beyond a certain limit \cite{Tremaine:we}. For the
Milky Way this means that the neutrino mass must be larger than
roughly 25 eV {\it if} neutrinos make up the dark matter. For
irregular dwarf galaxies this limit increases to 100-300 eV
\cite{Madsen:mz,salucci}, and means that standard model neutrinos
cannot make up a dominant fraction of the dark matter. This bound
is generally known as the Tremaine-Gunn bound.

Note that this phase space argument is a purely classical
argument, it is not related to the Pauli blocking principle for
fermions (although, by using the Pauli principle $f_\nu \leq 1$
one would arrive at a similar, but slightly weaker limit for
neutrinos). In fact the Tremaine-Gunn bound works even for bosons
if applied in a statistical sense \cite{Madsen:mz}, because even
though there is no upper bound on the fine grained distribution
function, only a very small number of particles reside at low
momenta (unless there is a condensate). Therefore, although the
exact value of the limit is model dependent, limit applies to any
species that was once in thermal equilibrium. A notable
counterexample is non-thermal axion dark matter which is produced
directly into a condensate.

It should also be noted that in the original work by Tremaine and Gunn the coarse grained distribution was assumed to be Maxwell-Boltzmann. However, this does not seriously influence the bound. Kull et al. \cite{kull} derived a more general bound
\begin{equation}
m \gg \left(\frac{\rho_\nu \sqrt{8 \pi}}{\sigma^3}\right)^{1/4},
\end{equation}
in which no specific assumption about the final distribution function is involved.

A very interesting direct example of the Tremaine-Gunn bound was
studied in \cite{Ringwald:2004np}. Here, neutrino clustering in
evolving cold dark matter halos was studied using the Boltzmann
equation. The problem was made tractable by assuming no
backreaction, i.e.\ that the CDM halos are not affected by
neutrinos. The results clearly show that there is a maximum in the
coarse grained distribution function of $1/2$, as expected. The
calculation was extended to bosons in \cite{Hannestad:2005bt},
where no such upper bound was found. In fact bosons can have an
average density several times higher than fermions with equal mass
inside dark matter halos, an effect which in principle might be
used to distinguish fermionic and bosonic hot dark matter.

Finally, the effect was studied in detailed $N$-body simulations in \cite{brandbyge4}. This point will be discussed in more detail below.

\subsection{Neutrino hot dark matter}

A much stronger upper bound on the neutrino mass than the one in
Eq.~(\ref{eq:mnu}) can be derived by noticing that the thermal
history of neutrinos is very different from that of a WIMP because
the neutrino only becomes non-relativistic very late.

The Boltzmann equation can generically be written as
\begin{equation}
L[f] = \frac{Df}{D\tau} = C[f],
\end{equation}
where $L[f]$ is the Liouville operator. The collision operator $C[f]$ on the right-hand side describes any
possible collisional interactions. For neutrinos $C[f]=0$ after neutrino decoupling at $T \sim 2-3$ MeV.

We then write the distribution function as
\begin{equation}
f(x^i,q,n_j,\eta) = f_0(q) [1+\Psi(x^i,q,n_j,\eta)], \label{eq:Psi}
\end{equation}
where $f_0(q)$ is the unperturbed distribution function. For a fermion which decouples while relativistic,
this distribution function is
\begin{equation}
f_0(q) = [\exp(q/T_0)+1]^{-1},
\end{equation}
where $T_0$ is the present-day temperature of the species.

In conformal Newtonian (longitudinal) gauge the Boltzmann equation for neutrinos can be written as an evolution equation
for $\Psi$ in $k$-space \cite{MB}
\begin{equation}
\frac{1}{f_0} L[f] = \frac{\partial \Psi}{\partial \tau} + ik \frac{q}{\epsilon} \mu \Psi + \frac{d \ln
f_0}{d \ln q} \left[\dot{\phi}-ik \frac{\epsilon}{q} \mu \psi \right] = 0,
\label{eq:boltzmann}
\end{equation}
where $\mu \equiv n^j \hat{k}_j$. $\psi$ and $\phi$ are the metric perturbations, defined from the perturbed
space-time metric in the conformal Newtonian gauge \cite{MB}
\begin{equation}
ds^2 = a^2(\eta) [-(1+2\psi)d\eta^2 + (1-2\phi)\delta_{ij} dx^i dx^j].
\end{equation}

The perturbation to the distribution function can be expanded as follows
\begin{equation}
\Psi = \sum_{l=0}^{\infty}(-i)^l (2l+1)\Psi_l P_l(\mu).
\end{equation}
One can then write the collisionless Boltzmann equation as a moment hierarchy for the $\Psi_l$'s by
performing the angular integration of $L[f]$
\begin{eqnarray}
\dot\Psi_0 & = & -k \frac{q}{\epsilon} \Psi_1 + \dot{\phi} \frac{d \ln f_0} {d \ln q}, \label{eq:psi0}\\
\dot\Psi_1 & = & k \frac{q}{3 \epsilon}(\Psi_0 - 2 \Psi_2) - \frac{\epsilon k}{3 q} \psi \frac{d \ln f_0}{d
\ln q}, \label{eq:psi1}\\ \dot\Psi_l & = & k \frac{q}{(2l+1)\epsilon}(l \Psi_{l-1} - (l+1)\Psi_{l+1}) \,\,\,
, \,\,\, l \geq 2.
\end{eqnarray}

\begin{figure}[htbp]
\begin{center}
\epsfig{file=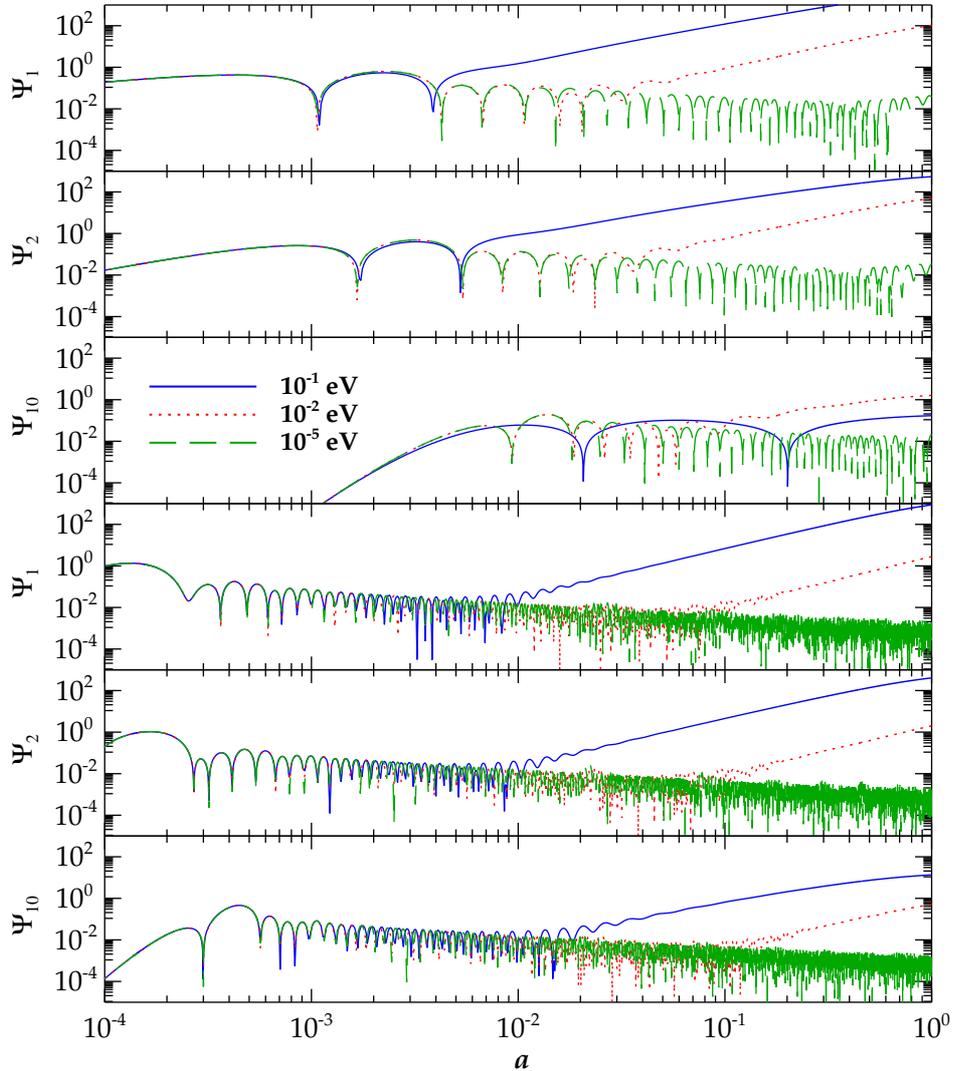,width=0.8\textwidth}
\end{center}
\bigskip
\caption{\label{fig:psi} $\Psi_l$'s for 3 neutrino masses with momentum $q/T_0=3$ as a function of the scale factor. The upper
   three panels are for $k = 0.01 \, h \, {\rm Mpc}^{-1}$ and the lower three panels for $k = 0.1 \, h \, {\rm
   Mpc}^{-1}$.}
\end{figure}

As long as neutrinos are relativistic the solution to this system of equations is damped for all modes
inside the horizon. Examples of the exact numerical solutions, calculated using CAMB \cite{CAMB} are presented
in Fig.~\ref{fig:psi}.
When the neutrino becomes non-relativistic the gravitational source term (the $k^2 \psi$ term) becomes important
and begins to feed the $\Psi$'s, starting with the lowest multipoles. The higher multipoles are affected slightly later.
Another important point is that $\Psi$ starts to grow later for the smaller $k$ value. This point will be discussed
below.

By integrating the neutrino perturbation over momentum
\begin{equation}
F_{\nu l} = \frac{\int dq q^2 \epsilon f_0(q) \Psi_l}{\int dq q^2 \epsilon f_0(q)},
\end{equation}
one finds a set of equations of which the first two orders are
\begin{eqnarray}
\dot \delta & = & -(1+w)(\theta - 3 \dot \phi) - 3 H (c_s^2-w)\delta, \\
\dot \theta & = & -H (1-3 w) \theta -\frac{\dot w}{1+w} \theta + \frac{c_s^2}{1+w} k^2 \delta - k^2 \sigma + k^2 \phi,
\end{eqnarray}
where $c_s^2 = \delta P/\delta \rho$, and $H = \dot a/a$. These equations can be recognised as the continuity and Euler
equations respectively.

Provided that the anisotropic stress term, $\sigma$ is negligible the hierarchy can be truncated after the $\theta$ term.
While this is true for a perfect fluid it is not so for massive neutrinos.
However, as discussed in detail in Ref.~\cite{kom102} the anisotropic stress for massive neutrinos in the non-relativistic regime can to a good
approximation be simply related to the $c_s^2$ term. The end result is a set of fluid equations for the massive neutrinos
which is approximately
\begin{eqnarray}
\dot \delta & = & - \theta, \\
\dot \theta & = & - H \theta - \left(\frac{3}{2}H^2 - \frac{5}{3} k^2 c_d^2\right) \delta,
\end{eqnarray}
where
\begin{equation}
c_d^2 = \frac{\int d^3 p f_0 v^2}{\int d^3 p f_0}.
\end{equation}

\begin{figure}[htbp]
\begin{center}
\epsfig{file=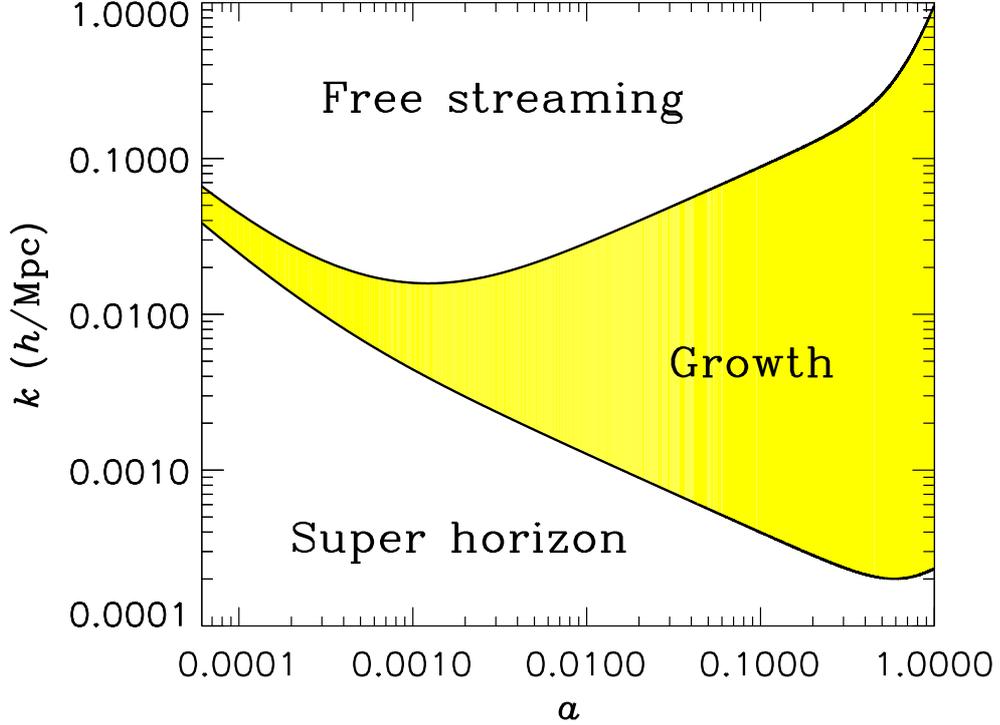,width=0.8\textwidth}
\end{center}
\bigskip
\caption{\label{fig:freestream} The upper black line is the neutrino free streaming scale, $k_{\rm FS}$, for a model with
$\sum m_\nu = 1.2$ eV. The lower black line is the comoving Hubble scale, $a H$. All modes in the shaded region can grow, whereas modes
above the shaded region are subject to free streaming suppression. Figure adapted from \cite{kom102}.}
\end{figure}

For non-relativistic neutrinos the anisotropic stress term acts like an additional pressure term which
prevents the growth of structure. Below a scale corresponding to
\begin{equation}
k_{\rm FS}^2 \sim \frac{9}{10} \frac{H^2}{c_d^2}
\end{equation}
structures cannot form.
For much smaller $k$ the solution corresponds to the growing mode of the CDM perturbations. In Fig.~\ref{fig:freestream} I have plotted the
free streaming scale as a function of $a$, as well as the comoving Hubble scale, $aH$. This clearly identifies the relatively narrow region in $k$ where modes can grow. When $a$ increases the neutrino velocity dispersion decreases and the region where growth is allowed becomes correspondingly larger. For a smaller neutrino mass the growing region would be smaller because $k_{\rm FS}$ starts to increase with $a$ only once neutrinos become non-relativistic. It is now also clear why the onset of growth of $\Psi$ seen in Fig.~\ref{fig:psi} happens later for larger $k$. The given
mode crosses into the region where growth is allowed at a later stage in cosmic evolution.

We are thus able to identify three different regions for $\tau$ and $k$:
\vspace*{0.2cm}\\
1) $\tau < \tau(T=m)$: Neutrinos are relativistic and there is no growth of structure,
\vspace*{0.2cm}\\
2) $\tau > \tau(T=m)$, $k < k_{\rm FS}$:
structures grow as for CDM,
\vspace*{0.2cm}\\
3) $\tau > \tau(T=m)$, $k > k_{\rm FS}$: Structures cannot form.
\vspace*{0.2cm}

\begin{figure}[htbp]
\begin{center}
\epsfig{file=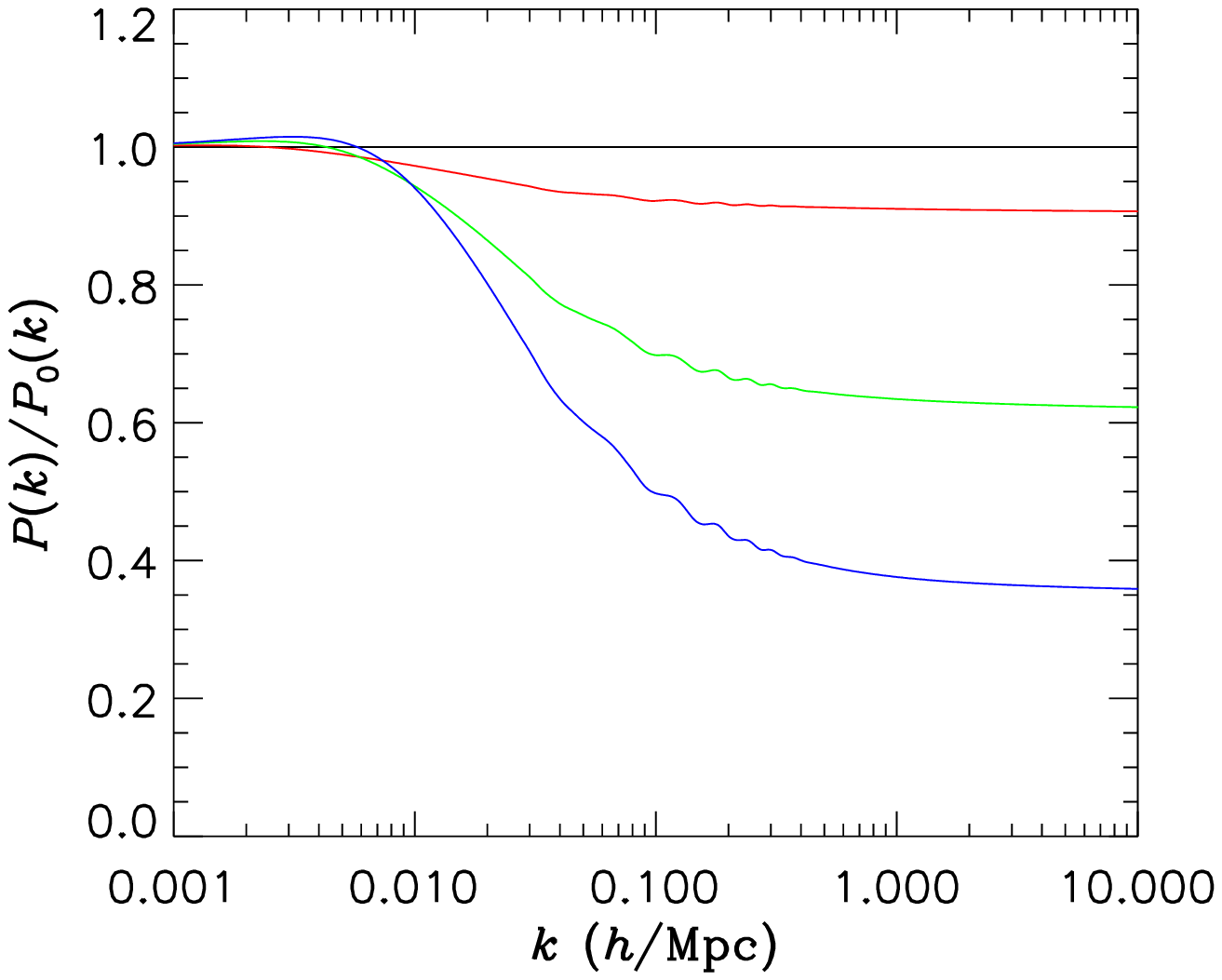,width=0.8\textwidth}
\end{center}
\bigskip
\caption{\label{fig:nutrans} The matter power spectrum divided by the $\Lambda$CDM matter power spectrum
for various different neutrino masses. The lines are for $\sum m_\nu = 0$, 0.15, 0.6, and 1.2 eV in descending order.}
\end{figure}

When measuring fluctuations it is customary to use the power
spectrum, $P(k,\tau)$, defined as
\begin{equation}
P(k,\tau) = |\delta|^2(\tau).
\end{equation}
The power spectrum can be decomposed into a primordial part,
$P_0(k)$, and a transfer function $T^2(k,\tau)$,
\begin{equation}
P(k,\tau) = P_0(k) T^2(k,\tau).
\end{equation}
The transfer function at a particular time is found by solving the
Boltzmann equation for $\delta(\tau)$.

In a universe with a mixture of neutrino hot dark matter and cold dark matter, the suppression of fluctuations on small scales can be found using an analytic approximation. On scales which are smaller than the free-streaming scale for neutrinos, the usual Meszaros equation for the evolution of CDM perturbations is given by
\begin{equation}
\ddot \delta + \frac{2}{\tau} \dot\delta - \frac{6}{\tau^2}(1-f_\nu) \delta = 0,
\end{equation}
i.e.\ the gravitational source term is modified because neutrinos do not contribute to the gravitational potential, only to the background evolution.
The normal growing solution $\delta \propto \tau^2$ is then modified to $\delta \propto \tau^{(\sqrt{1+24 (1-f_\nu)}-1)/2}$, and in the limit of $f_\nu \ll 1$ the power spectrum is modified roughly by \cite{Lesgourgues:2006nd}
\begin{equation}
\frac{\Delta P}{P} \sim -8 f_\nu.
\end{equation}
The same suppression factor is found from an exact numerical solution to the Boltzmann hierarchy \cite{Hu:1997mj}.
In Fig.~\ref{fig:nutrans} I have plotted matter power spectra for
various different neutrino masses in a flat $\Lambda$CDM universe
$(\Omega_m+\Omega_\nu+\Omega_\Lambda=1)$. The parameters used were
$\Omega_b = 0.04$, $\Omega_{\rm CDM} = 0.26 - \Omega_\nu$,
$\Omega_\Lambda = 0.7$, $h = 0.7$, and $n=1$ and the calculations were done with the CAMB \cite{CAMB} Boltzmann solver.

The effect of massive neutrinos on structure formation only
applies to the scales below the free-streaming length. For
neutrinos with masses of several eV the free-streaming scale is
smaller than the scales which can be probed using present CMB data
and therefore the power spectrum suppression can be seen only in
large scale structure data. On the other hand, neutrinos of sub-eV
mass behave almost like a relativistic neutrino species for CMB
considerations. The main effect of a small neutrino mass on the
CMB is that it leads to an enhanced early ISW effect. The reason
is that the ratio of radiation to matter at recombination becomes
larger because a sub-eV neutrino is still relativistic or
semi-relativistic at recombination.

\subsection{Non-linear evolution}

As was seen in the previous section the effect of neutrinos on the matter power spectrum in the linear regime can be reasonably well approximated by
\begin{equation}
P_m = \cases{P_m(\Omega_m = \Omega_c + \Omega_\nu) & for $k \ll k_{\rm FS}$ \cr
(1-8 f_\nu) P_m(\Omega_m = \Omega_c + \Omega_\nu) & for $k \gg k_{\rm FS}$},
\end{equation}
with a smooth transition stretching over roughly 2 decades in $k$.

However, at $z = 0$ non-linear corrections become important already at $k \sim 0.1 \, h$/Mpc. These non-linear corrections can be described either semi-analytically using the halo-model formalism, or using an extension of perturbation theory.
Analytic calculations of the matter power spectrum with massive neutrinos included have been carried out in a number of recent studies
\cite{Saito:2008bp,Wong:2008ws,Lesgourgues:2009am,Saito:2009ah}.

Ultimately, however, the validity of these approximations must be tested against N-body simulations. However,
massive neutrinos pose a serious problem for N-body simulations since they have very high thermal velocities. For realistic masses most neutrinos will have much higher thermal velocities than the average gravitational streaming velocities, meaning that most neutrinos do not cluster in bound halos. Simply treating neutrinos as particles is difficult because the thermal motion introduces noise which completely dominates the behaviour of neutrinos on small scales. However, it has been shown that the matter power spectrum can be computed at the 1\% level of precision with neutrinos included, using N-body techniques \cite{Brandbyge1,Brandbyge2,Brandbyge3,brandbyge4}.
These simulations have shown a very interesting connection between thermal neutrino motion and gravitational streaming motion. In general neutrinos cause a suppression of matter fluctuations which is larger than the linear theory result, particularly on scales corresponding to the free-streaming scale at a {\it given} redshift, i.e.\ on scales where the virial velocity of a halo is comparable to the thermal velocity of a typical neutrino.
The maximal suppression is roughly $-9.6 f_\nu$, peaked at scales around $k \sim 0.5-1 \, h$/Mpc. Provided that other aspects of the power spectrum can be modeled sufficiently precisely this feature is a smoking gun signature for the presence of massive neutrinos. The presence of this neutrino induced depression in the power spectrum cannot be mimicked by other types of dark matter or dark energy, and furthermore it is sufficiently far removed from the baryon acoustic peak (at $k \sim 0.15 \, h$/Mpc) to be uniquely identified.
It should also be noted that a peak suppression of $\sim -9.6 f_\nu$ was also found in a recent study by Viel and Springel \cite{Viel:2010bn} which focussed mainly on simulations of the Lyman-$\alpha$ forest.
Finally it should be noted that calculations using various extensions of linear perturbation theory, as described in \cite{Saito:2008bp,Wong:2008ws,Lesgourgues:2009am,Saito:2009ah} show good agreement with simulations at high $z$ and for relatively small $k$, i.e.\ the regime where non-linear corrections are mild, but not for larger values of $k$, i.e.\ $k \gtrsim 0.3 \, h$/Mpc.

\begin{figure}[htbp]
\begin{center}
\epsfig{file=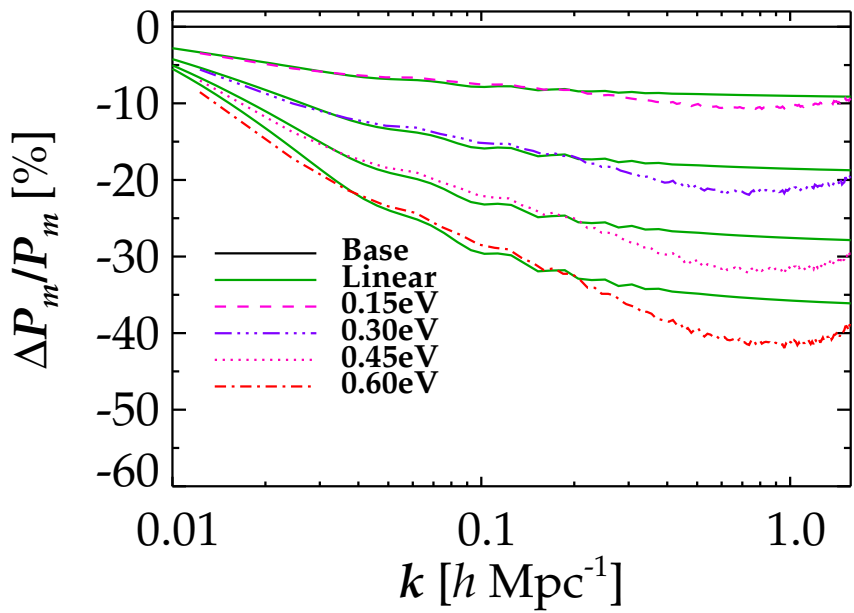,width=0.8\textwidth}
\end{center}
\bigskip
\caption{\label{fig:nutrans2} The matter power spectrum divided by the $\Lambda$CDM matter power spectrum
for various different neutrino masses. The lines are for $\sum m_\nu = 0$, 0.15, 0.6, and 1.2 eV in descending order. The green lines are
the linear theory predictions (figure reproduced from \cite{hb10}).}
\end{figure}

\subsection{Other aspects of non-linear structure formation}

The matter power spectrum is not the only interesting large scale structure observable. Other probes which will be of importance in the next decade are for example weak gravitational lensing and the measurement of galaxy clusters. In \cite{brandbyge4} properties of dark matter halos in cosmologies with massive neutrinos were studied and found to be significantly different from standard $\Lambda$CDM halos. In general halos of a given mass form later and are less concentrated in such models, and the number of halos of a given mass is also changed. Part of the effect comes from the different linear theory initial condition, but there is also a substantial effect from non-linear evolution.
In \cite{Wang:2005vr} possible neutrino mass constraints from a future cluster survey were studied in a Fisher matrix analysis and using the Jenkins {\it et al.} \cite{Jenkins:2000bv} prescription for propagating the linear theory transfer function to the halo mass function. It was found that some of the cluster surveys which will be carried out in the next decade will allow for 0.05 eV sensitivity to neutrino mass.

\subsubsection{The halo mass function}
\label{sec:hmf}

In \cite{brandbyge4} detailed studies were performed on how clusters form in cosmologies with massive neutrinos. The numerical set-up was based on the
code described in \cite{Brandbyge3} and from the N-body simulations, halos were identified using the AMIGA halo finder
\cite{Knollmann:2009pb}.

\begin{figure}
    \noindent
     \begin{center}
           \includegraphics[width=0.7\linewidth]{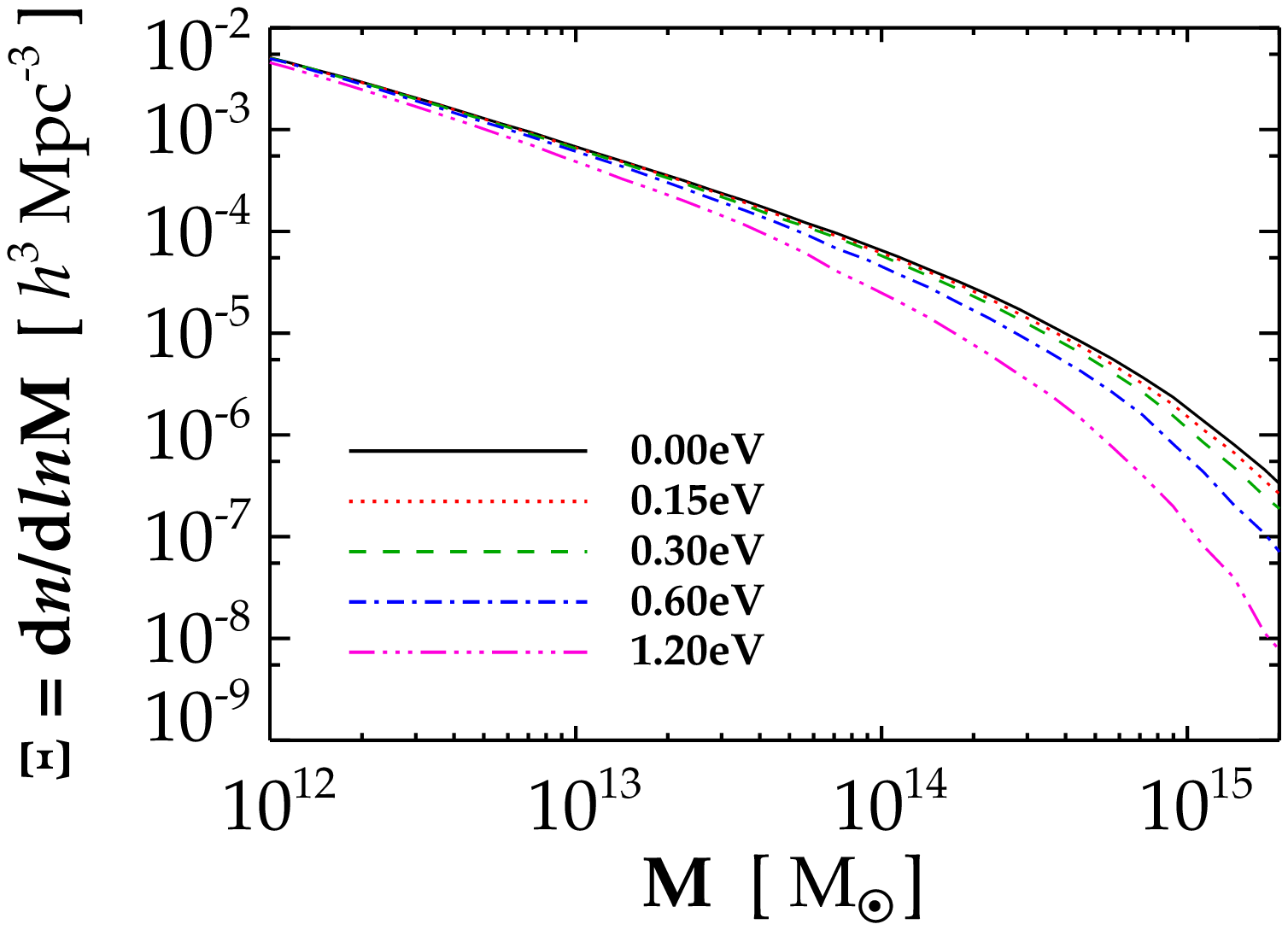}
     \end{center}
     \vspace*{-1.2cm}

     \begin{center}
           \includegraphics[width=0.7\linewidth]{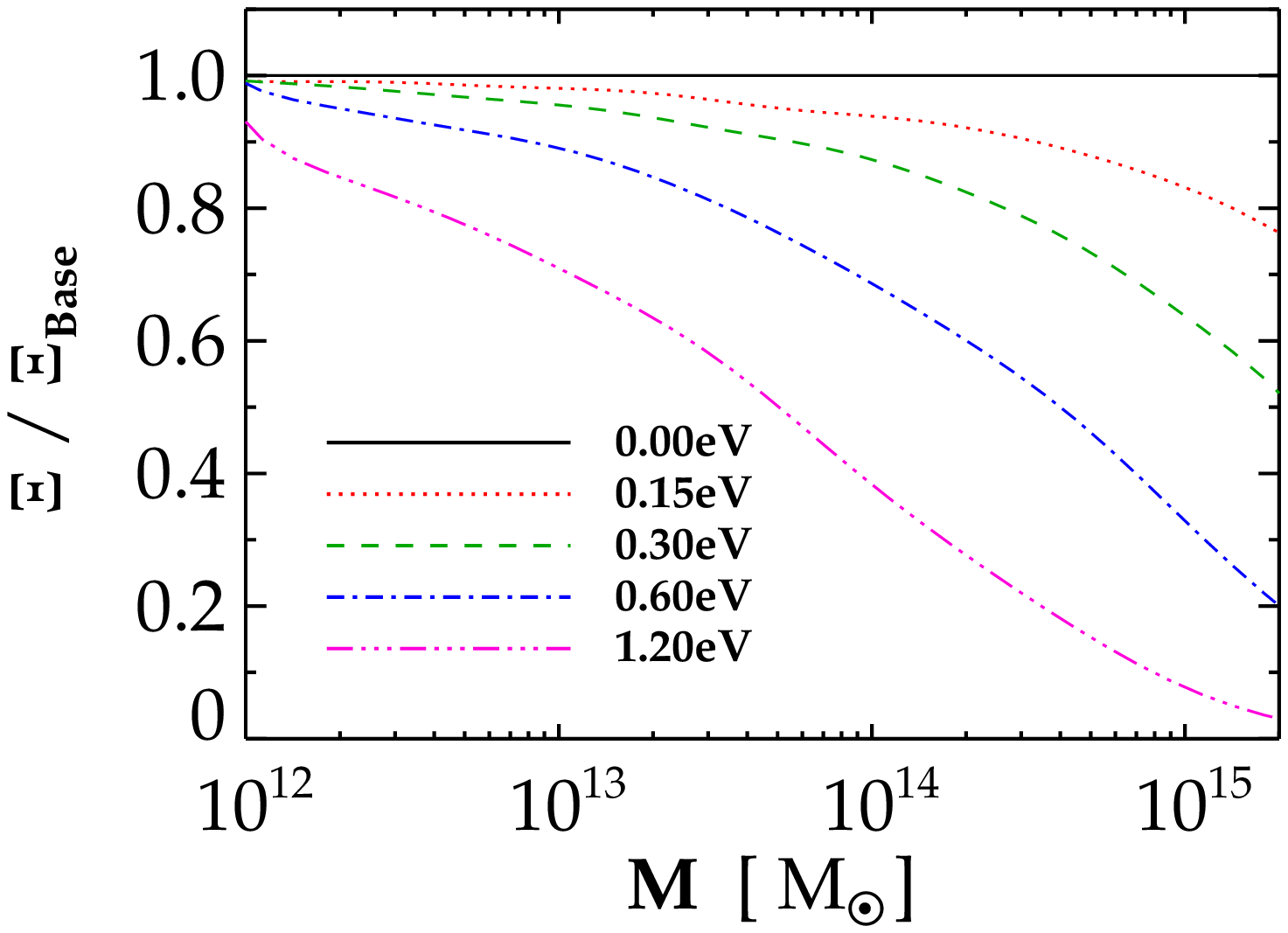}
     \end{center}
     \vspace*{-0.5cm}

   \caption{Absolute and relative halo mass functions for 5 different neutrino cosmologies. The halo mass functions have been splined and smoothed together to obtain sufficient accuracy in the halo mass range $10^{12}$ to $10^{15}$ ${\rm M}_\odot$ (figure reproduced from \cite{brandbyge4}).}
   \label{fig:halo_mass_function}
\end{figure}

\begin{figure}
    \noindent
     \begin{center}
           \includegraphics[width=0.7\linewidth]{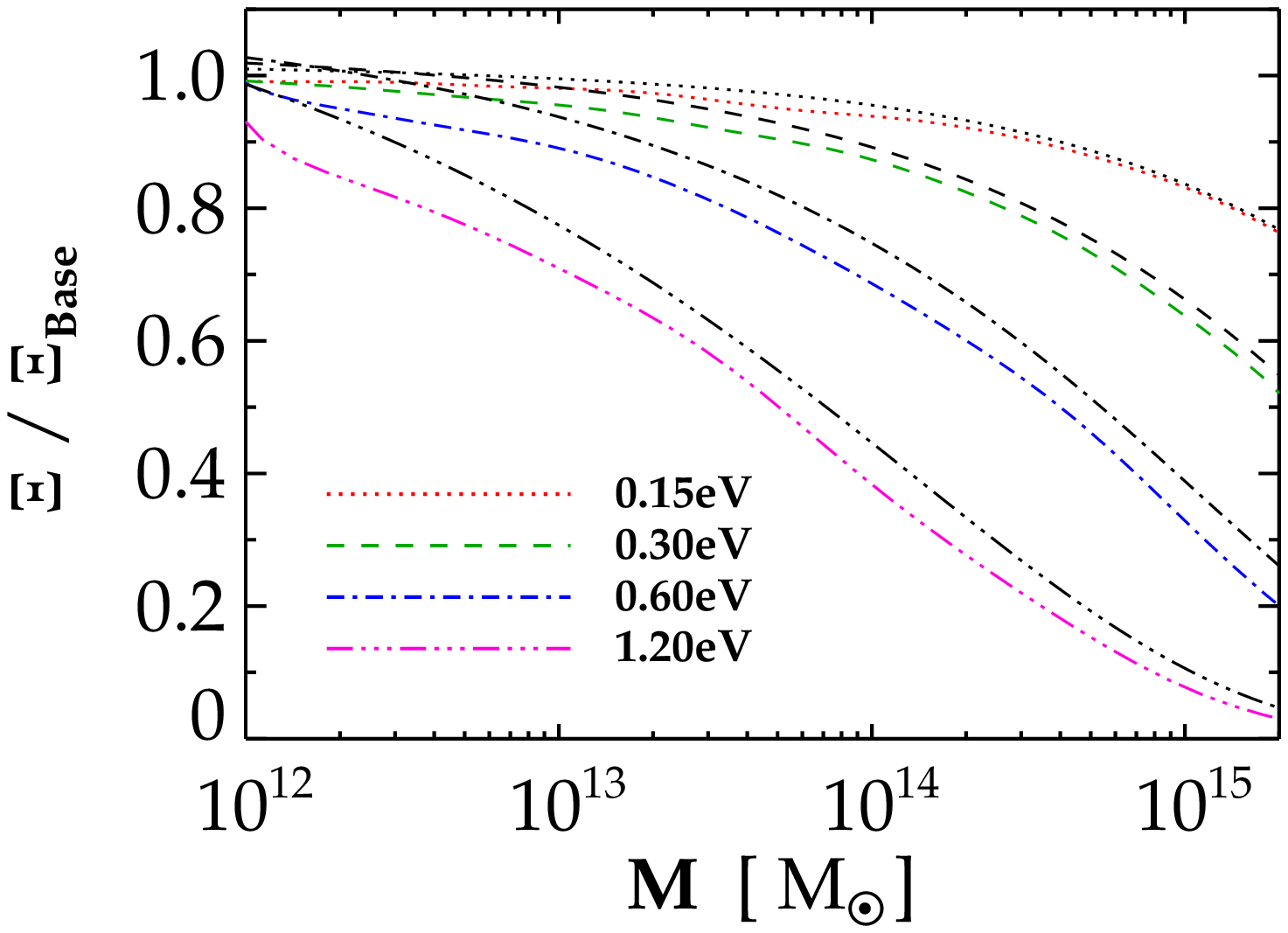}
     \end{center}
     \vspace*{-1.2cm}

     \begin{center}
           \includegraphics[width=0.7\linewidth]{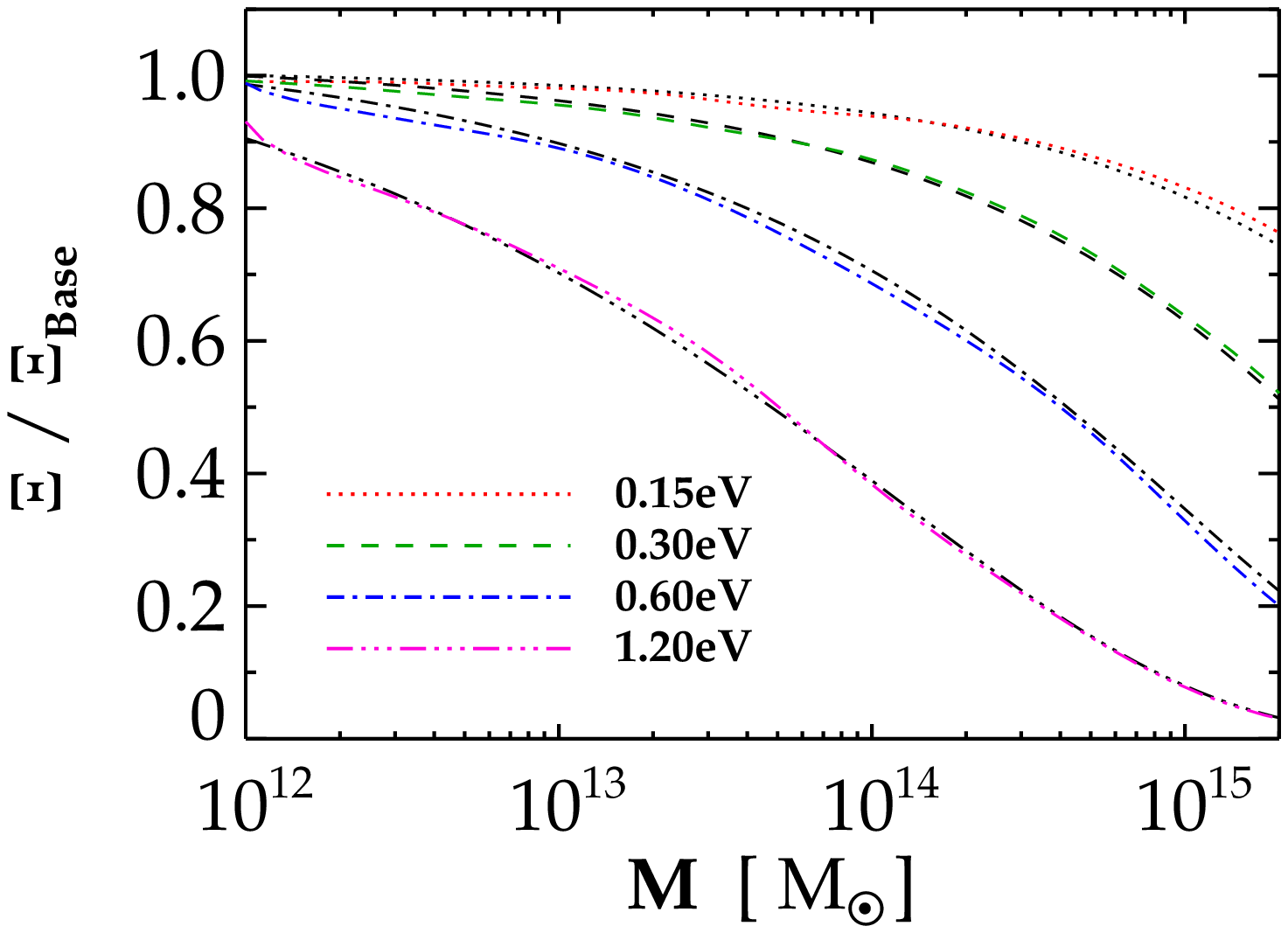}
     \end{center}
     \vspace*{-1.2cm}

     \begin{center}
           \includegraphics[width=0.7\linewidth]{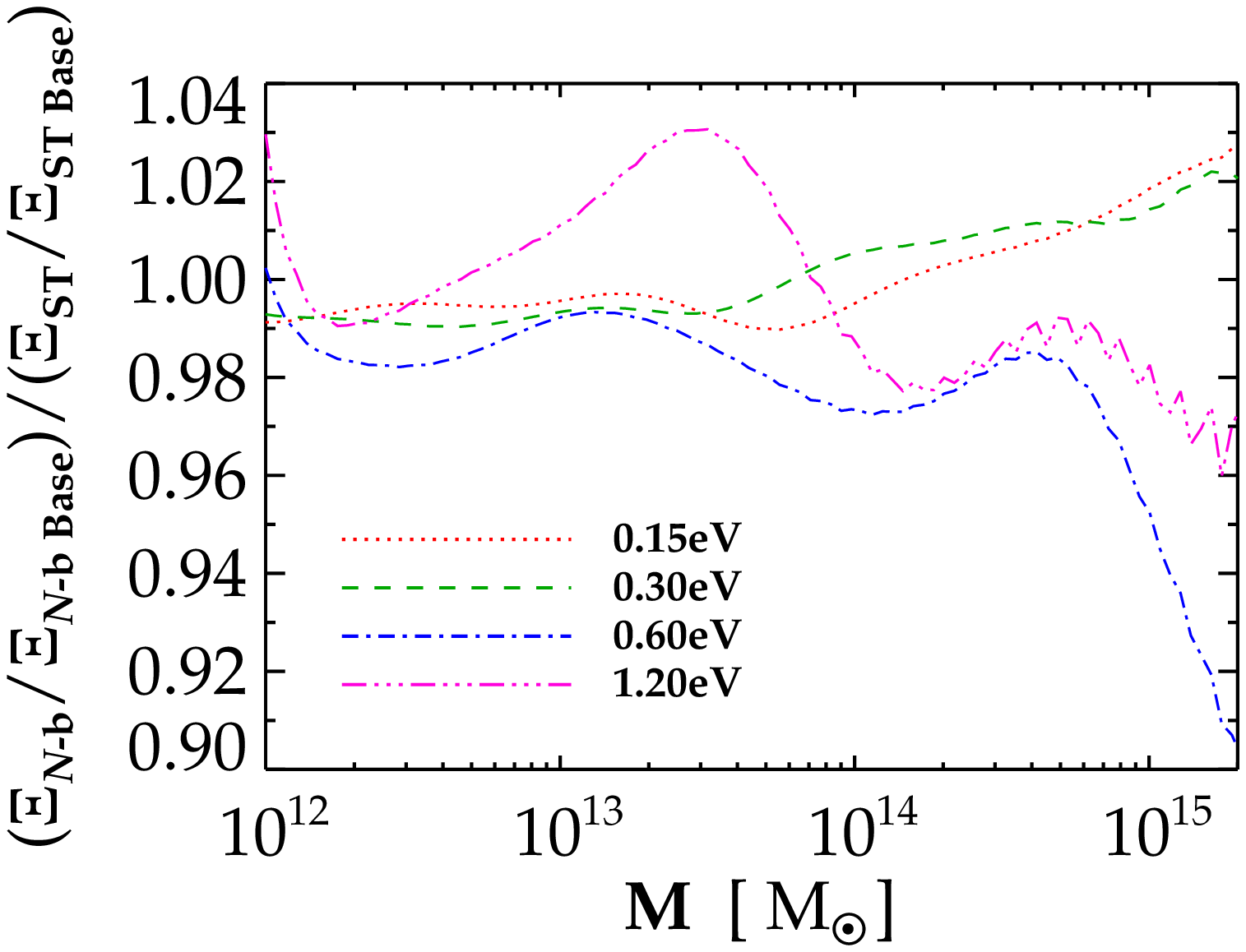}
     \end{center}
     \vspace*{-0.5cm}

   \caption{Relative halo mass functions for different neutrino cosmologies compared with the predictions from the Sheth-Tormen formulae (black lines). Top: With $\Omega_m = \Omega_c + \Omega_b + \Omega_\nu$ in the ST formulae. Middle: With $\Omega_c + \Omega_b$ used instead of $\Omega_m$ in the ST formulae. Bottom: Differences between the $N$-body and the ST predictions (figure reproduced from \cite{brandbyge4}).}
   \label{fig:shethtormen}
\end{figure}

From the list of halos the number density of halos per mass interval, i.e.\ the halo mass function (HMF) can be constructed.
In Fig.~\ref{fig:halo_mass_function} the HMFs for cosmologies with different neutrino masses are shown. As expected the HMF is more suppressed in cosmologies with a larger neutrino mass. The suppression is largest for the heaviest, late forming halos.

In Fig.~\ref{fig:shethtormen} the HMFs calculated from $N$-body simulations are compared with HMFs from the Sheth-Tormen (ST) semi-analytic formulae \cite{Sheth:2001dp}.
The ST fit is based on the fact that, as first pointed out by Press and Schechter \cite{Press:1973iz}, the HMF can be written as
\begin{equation}
\frac{M dM}{\bar \rho}\frac{dn(M,z)}{dM} = \nu f(\nu) \frac{d\nu}{\nu},
\end{equation}
with $\nu \equiv [\delta_{\rm sc}(z)/\sigma(M)]^2$, where $\delta_{\rm sc}(z)=1.686$ is the overdensity required for spherical collapse at $z$, and $\bar \rho = \Omega_m \rho_c$.
$dn(M,z)$ is the number density of halos in the mass interval $M$ to $M + dM$.
The variance of the linear theory density field, $\sigma^2(M)$, is given by
\begin{equation}
\sigma^2(M) = \int \frac{dk}{k} \frac{k^3 P_{\rm lin}(k)}{2 \pi^2} |W(kR)|^2,
\end{equation}
where $P_{\rm lin}(k)$ is the linear theory matter power spectrum, and the Top-Hat window function is given by $W(x) = (3/x^2)(\sin x - x \cos x)$ with $R = (3M/4\pi \bar \rho)^{1/3}$.

The ST fit to $\nu f(\nu)$ is
\begin{equation}
\nu f(\nu) = A \left(1+\frac{1}{\nu'^p}\right)\left(\frac{\nu'}{2}\right)^{1/2} \frac{e^{-\nu'/2}}{\sqrt{\pi}},
\end{equation}
with $\nu' = 0.707 \nu$ and $p=0.3$. $A=0.3222$ is determined from the integral constraint $\int f(\nu) d\nu = 1$.

The upper panel in Fig.~\ref{fig:shethtormen} shows that the agreement is poor if $\Omega_m = \Omega_c + \Omega_b + \Omega_\nu$ is used in the ST formalism. However, this is due to a wrong definition of the halo mass: Even for the very largest cluster halos the neutrino component contributes very little to the halo mass. In reality, the mass inside the collapsing region should be calculated using $\Omega_c + \Omega_b$, not $\Omega_m$. This amounts to neglecting the weakly clustering neutrino component when calculating the halo mass.
The two lower panels in Fig.~\ref{fig:shethtormen} shows the same ST fit, but using $\Omega_c + \Omega_b$ instead of $\Omega_m$. In this case the ST HMFs provide an excellent fit to the
relative change to the HMF caused by neutrinos. As the figure at the bottom clearly demonstrates, the agreement is better than $\sim 3\%$ at halo mass scales where the $N$-body HMFs are accurate.
Although the absolute HMFs, even for CDM simulations, do not match the ST HMFs more precisely than at the $\sim 10$\% level, the {\it relative} change from adding neutrinos can be calculated significantly more accurately.

\subsubsection{Individual neutrino halos}

\begin{figure}
     \noindent
     \vspace*{-1cm}
     \begin{center}
           \includegraphics[width=0.7\linewidth]{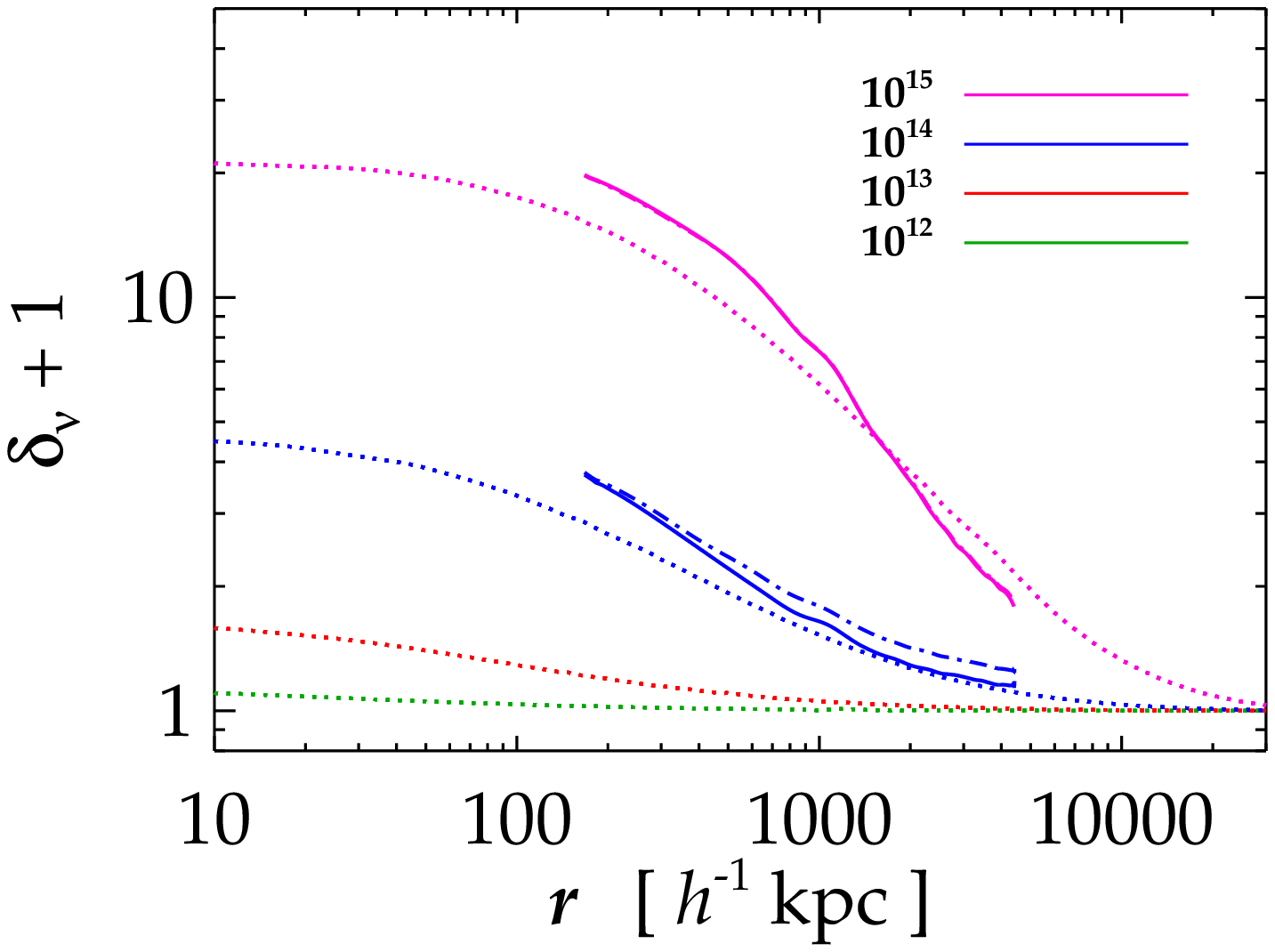}
     \end{center}
     \vspace*{-1cm}

     \begin{center}
           \includegraphics[width=0.7\linewidth]{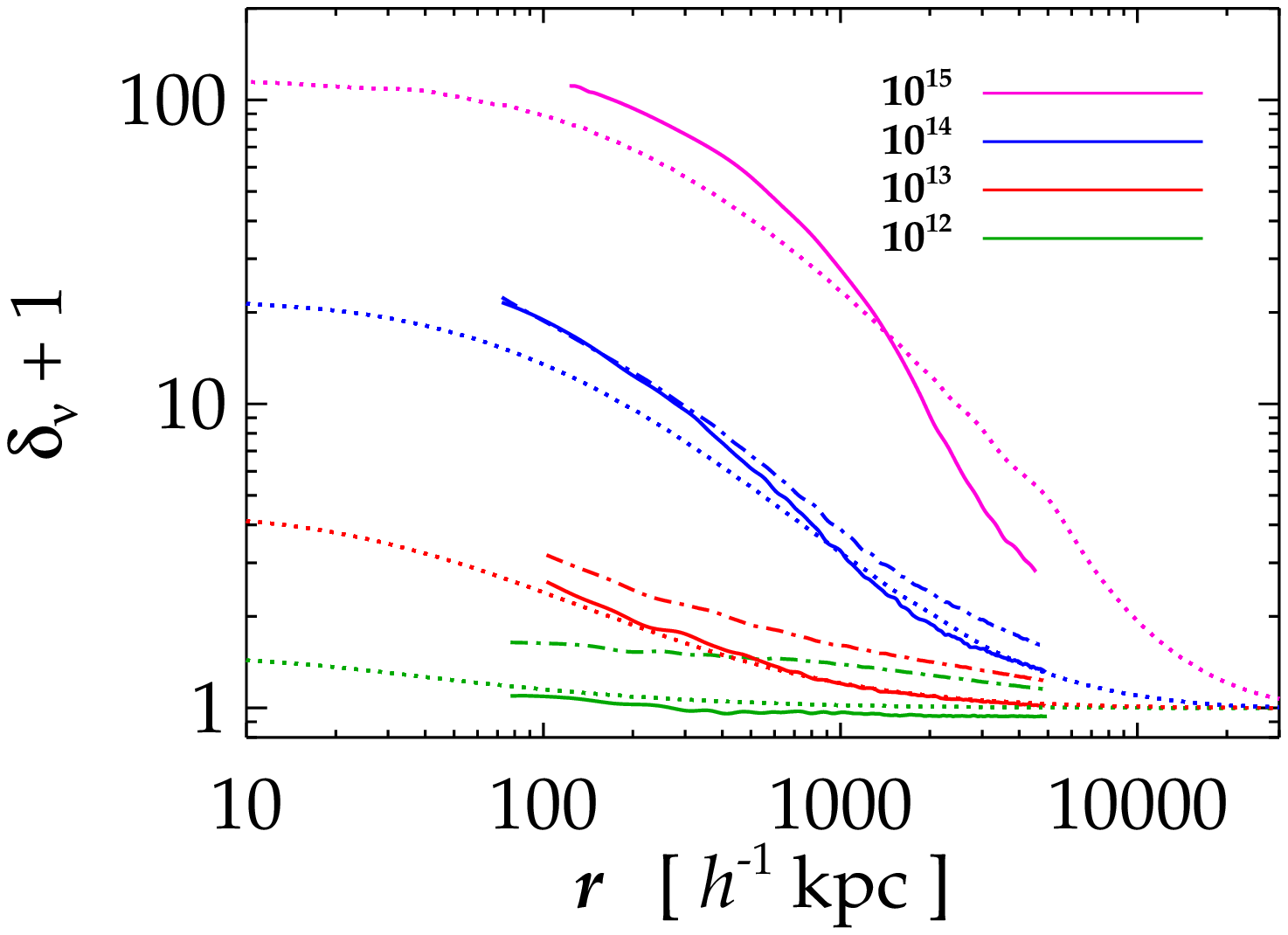}
     \end{center}
     \vspace*{-1cm}

     \begin{center}
           \includegraphics[width=0.7\linewidth]{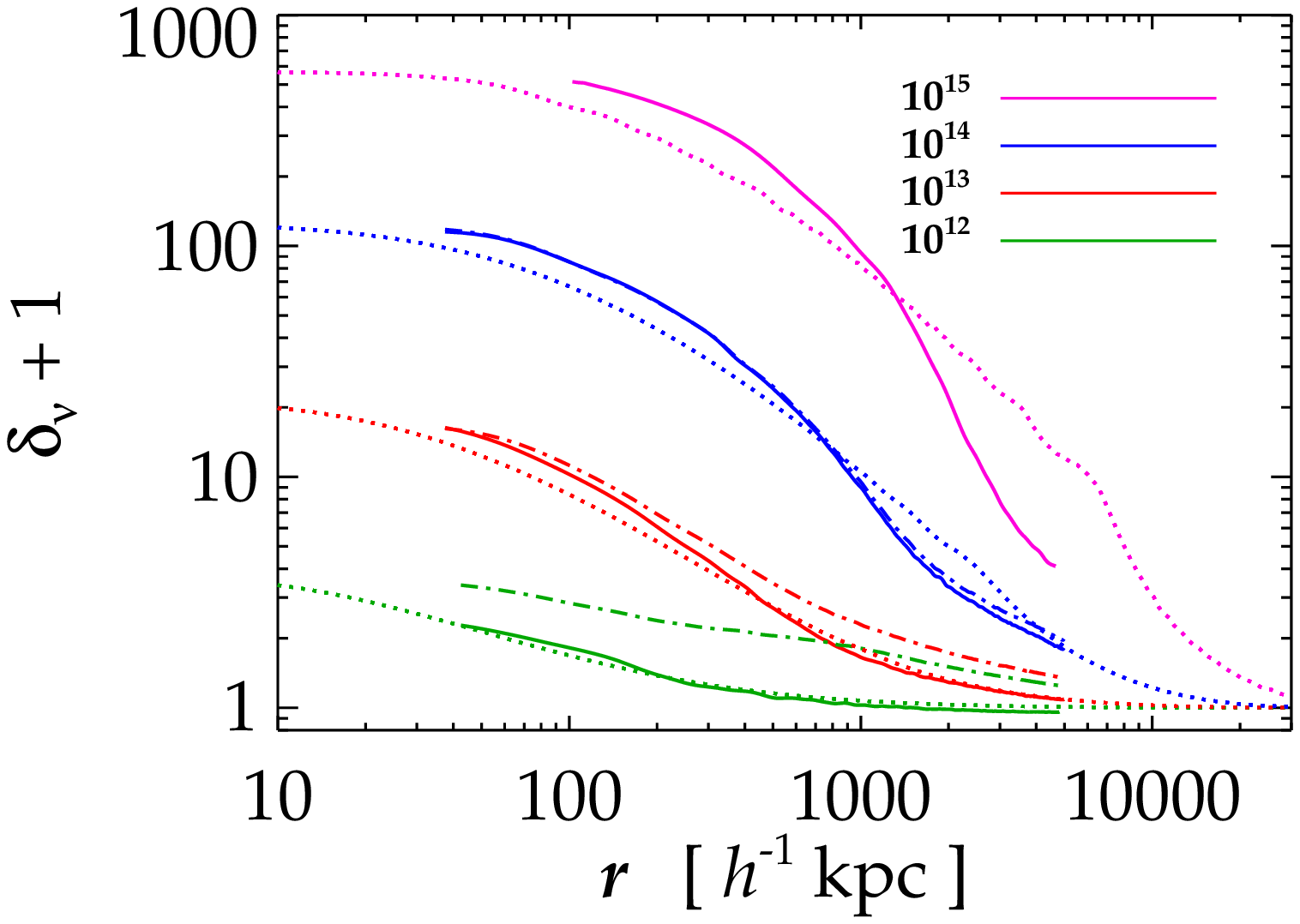}
     \end{center}
     \vspace*{-0.5cm}
     \caption{Neutrino halo profiles for $\sum m_\nu = 0.3 \, {\rm eV}$ (top),  $\sum m_\nu = 0.6 \, {\rm eV}$ (middle) and $\sum m_\nu = 1.2 \, {\rm eV}$ (bottom) for halo masses of $10^{12}$,  $10^{13}$, $10^{14}$ and $10^{15} \, {\rm M}_\odot$. Profiles are calculated with the $N$-one-body method (dotted) and the $N$-body method with a halo isolation criterion (solid) and without (dot-dashed) (figure reproduced from \cite{brandbyge4}).}
   \label{fig:nu_profiles}
\end{figure}

Detailed N-body simulations can also be used to probe the density profiles of individual halos. In \cite{brandbyge4} the overdensity of neutrinos in halos of varying size was studied for a variety of different neutrino masses. The results of these simulations are shown in Fig.~\ref{fig:nu_profiles}. The figure shows three different types of curves: One for all halos of a given total mass, one with only isolated halos, i.e.\ halos which are not subhalos of a larger halo, and one which is calculated using the N-1-body method \cite{Ringwald:2004np} (see \cite{brandbyge4} for details).

The gravitational effect of a host halo is relatively much more important for neutrinos than for the CDM component: Due to free-streaming neutrinos will almost completely stream out of small halos ($\simeq 10^{12}{\rm M}_\odot$), and any measured value $\delta_\nu > 0$ will be caused by the host halo. The radial profile of $\delta_\nu$ will therefore be a superposition of a dominant flat profile from the host halo on top of a sub-dominant contribution from the $\simeq 10^{12}{\rm M}_\odot$ halo itself. This fact can be seen in Fig.~\ref{fig:nu_profiles}.

\begin{figure}
    \includegraphics[width=0.45\linewidth, angle = 90]{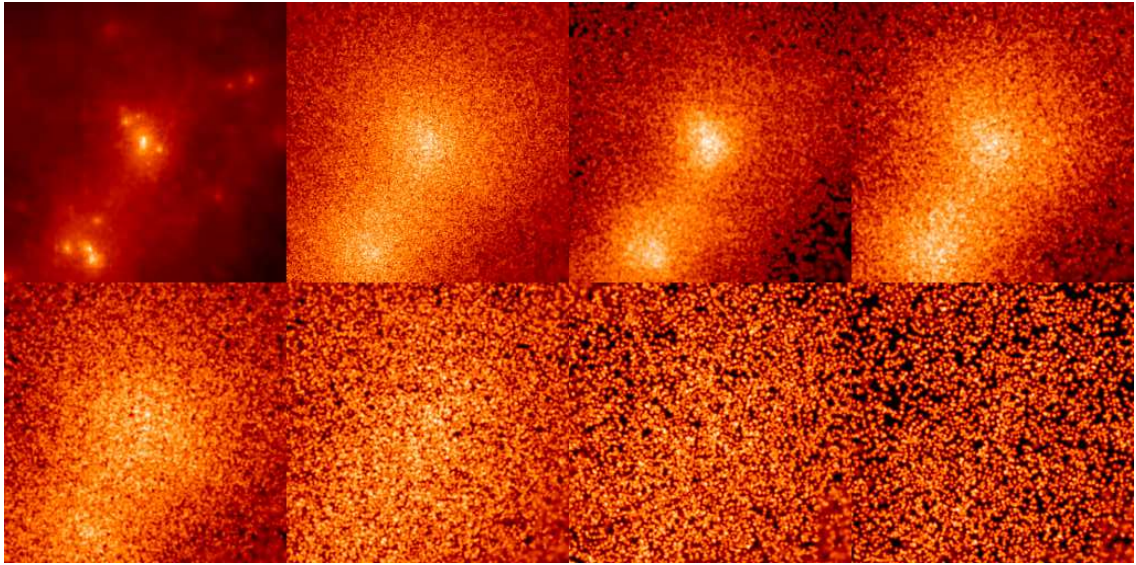}\vspace*{0.1cm}
    \caption{CDM and $\sum m_\nu = 1.2\, {\rm eV}$ neutrino distributions for a halo mass of $\simeq 5\cdot 10^{14}{\rm M}_\odot$. Dimensions in each image is 5 $h^{-1} \, {\rm Mpc}$. The images correspond to CDM, total neutrino, and $q/T = 1$ to 6 from top-left to bottom-right. Individual neutrino $N$-body particles can be identified (figure reproduced from \cite{brandbyge4}).}
   \label{fig:image1}
\end{figure}

In the same study it was also shown that the Tremaine-Gunn bound for neutrinos is in practise never saturated because neutrinos only start clustering very late in the evolution of the universe. In practise even the central density of neutrinos is significantly lower than what is estimated from the TG bound for a given neutrino mass.
Finally, in Fig.~\ref{fig:image1} I show an example of a neutrino halo in a heavy host halo of $5\cdot 10^{14}{\rm M}_\odot$.

\section{Current constraints on the neutrino mass}

\subsection{Parameter estimation methodology}

Because massive neutrinos affect structure formation it is
possible to constrain their mass using a combination of
cosmological data. The standard approach to cosmological parameter
estimation is to use Bayesian statistics which provides a very
simple method for incorporating prior information on parameters
from other sources. Using the prior probability distribution it is
then possible to calculate the posterior distribution and from
that to derive confidence limits on parameters. There are standard
packages such as CosmoMC \cite{cosmomc}, a likelihood calculator
based on the Markov Chain Monte Carlo method
\cite{Christensen:2001gj,CAMB}, available for this
purpose.

Although it is standard practise in cosmology to analyze data using a Bayesian framework
there are examples of frequentist analyses. In most cases there is little difference in the inferred
parameter ranges as long as the given parameters are well constrained by the data.

Independent of the statistical method used results will in general
depend on the number of parameters used to fit the data. Because
of parameter degeneracies bounds on a given parameter will in
general get weaker if more parameters are used in the fit.
However, the question remains as to how many parameters should
plausibly be included.
Particularly with early data it was always a problem that changes in parameters like the neutrino
mass could be mimicked by a combination of changes in other parameters, i.e.\ there were severe
parameter degeneracies in the data.
In such cases the actual constraint on a given parameter can depend strongly on the chosen model space.
For example the constraint on neutrino mass from CMB and large scale structure data would degrade by a factor
of two or more when the equation of state of dark energy was allowed to be different from -1.

With current data this is less of an issue simply because better data causes many parameter degeneracies to
be broken. In some cases this is true for just a single set of data, but in other cases the breaking of degeneracies
arises from the combination of two or more data sets.

\subsection{Current bounds}

Unlike the first WMAP data, the WMAP-7 data release on its own provides a quite stringent constraint on the sum of
neutrino masses of $\sum m_\nu < 1.3$ eV at $95 \%$ c.l.~\cite{kom10} (down from $\sim 2.1$ eV in the first releases
\cite{Ichikawa:2004zi}).
However, this remains true only within the framework of the standard $\Lambda$CDM + $m_\nu$ model.
For example, even if a prior on $H_0$ corresponding to the latest HST result \cite{Riess:2009pu} is added to the CMB data
the bound worsens to $\sum m_\nu < 2.1$ eV at $95 \%$ c.l. if $N_\nu \neq 3.04$ is allowed and further to 2.6 if the dark
energy equation of state, $w$, is allowed to be different from -1 \cite{Hamann:2010pw}.

Including large scale structure data the bound can be improved significantly. At present, by far the best large scale
structure data comes from  the Sloan Digital Sky Survey, for example in the form of the halo power spectrum (HPS) from the
SDSS-LRG DR7 sample presented in \cite{Reid:2009xm}, or from the BAO feature determined from the full SDSS-DR7 data set
\cite{Percival:2009xn}.

When combined with CMB data the HPS provides an upper bound of $0.61$ eV in the minimal model (going down to 0.44 eV if the HST prior
is imposed), and 1.16 eV in the extended
$\Lambda$CDM+$m_\nu$+$N_\nu$+$w$ model. The corresponding CMB plus BAO constraint is $0.85$ eV for the minimal model and 1.4 eV
for the extended model \cite{Hamann:2010pw}.
Note also the recent analysis performed in \cite{thomas2009} of the photometric SDSS MegaZ LRG DR7 catalogue which gives a very
stringent bound of $0.68$ eV when combined with other data. Ref.~\cite{swansson} contains an interesting analysis of the robustness of the mass bound against astrophysical uncertainties.

Another similar SDSS related probe is the SDSS MaxBCG catalog, used in \cite{Reid:2009nq}. Effectively this provides a bound on
the small scale normalisation of fluctuation power, and yields an upper bound on $0.4$ eV for the minimal model.
Another probe which in principle is very sensitive to the small scale power is the Lyman-$\alpha$ forest. The most recent
and sophisticated analysis of Lyman-$\alpha$ data \cite{Viel:2010bn} gives an upper bound of 0.9 eV for the sum of neutrino masses.

In conclusion the current bound on the sum of neutrino masses can
be in the range between 0.3 and more than 2 eV, depending on the data and
parameters used. In general:
\vspace*{0.2cm}\\
a) The bound on $\sum m_\nu$ for the minimal $\Lambda$CDM plus neutrino mass (8
parameters in total) is in the 0.4 eV range if CMB and LSS data is used.
\vspace*{0.2cm}\\
b) This bound can be relaxed somewhat when more
parameters, such as $w$ and $N_\nu$, are included. In the most conservative case the bound is above
2.5 eV if only CMB data is used.
\vspace*{0.2cm}\\
c) When CMB data is combined with LSS data in the linear or almost linear regime, combined with a
prior on the Hubble parameter the upper bound is robustly below 1 eV. This is true even for extended
models.
\vspace*{0.2cm}

\begin{figure}
\hspace{25mm}
\includegraphics[height=12.5cm,angle=-90]{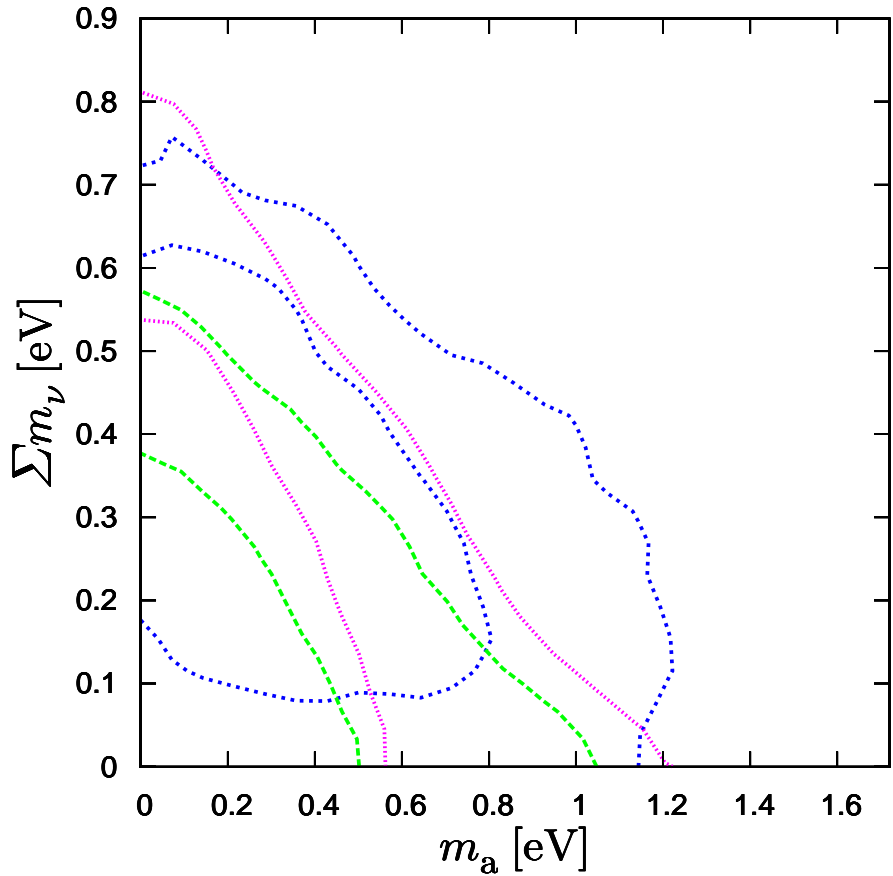}
\caption{2D marginal 68\% and 95\% contours in the $\sum m_\nu$-$m_a$
  plane.  The blue lines correspond to our old results using WMAP-5,
the magenta lines our new results using CMB+HPS, and the green lines using CMB+HPS+HST (figure reproduced from \cite{Hannestad:2010yi}).
\label{fig:mnu_ma}}
\end{figure}

Here it should perhaps also be noted that the bound on neutrino mass from cosmic structure formation applies to any
other, hypothetical particle species which decouples while still relativistic. This could for example be low mass sterile
neutrinos, as discussed for example in \cite{Dodelson:2005tp,acero}. It could also be relatively high mass axions which decouple
after the QCD phase transition \cite{Hannestad:2003ye,Hannestad:2010yi}. While there are some quantitative differences, the
qualitative argument is the same for all such cases.
In Fig.~\ref{fig:mnu_ma} I show one such example of a combined analysis of neutrinos and axions.

\section{Is there evidence for extra neutrino density?}

The WMAP data have consistently pointed to a value of $N_\nu$ larger than 3 as the best fit, and this is particularly true for the latest WMAP-7 data release \cite{kom10}. With the WMAP-7 data plus the SDSS-LRG DR7 data it was found by the WMAP team that $N_\nu = 4.25_{-0.80}^{+0.76}$ \cite{kom10}, and similar results have been obtained in other similar analyses. While not statistically significant at more than $2\sigma$ it is nevertheless suggestive, and also leads to the question of what this extra energy density could be.

In terms of neutrinos the simplest possibility by far is one or more light sterile states mixing with the active flavours. Such neutrinos can be either partly or completely thermalised via mixing in the early universe, with their final abundance depending strongly on the mass difference and mixing angle with the active species.

For a fully or almost thermalised sterile state the usual cosmological mass bound applies and means that
it cannot have a mass above 0.5-1 eV in order not to be in conflict with structure formation constraints \cite{Dodelson:2005tp,acero}.
However, for small mixing angles full thermalisation does not happen and the mass bound is significantly relaxed \cite{acero}. This would for example be the case for keV sterile neutrino warm dark matter. The most recent study of light sterile neutrinos in cosmology is the one performed in \cite{hamann2010}.

Another possibility which is perhaps slightly less natural from a model building perspective, but not excluded, is the presence of large neutrino chemical potentials. In terms of energy density and non-zero chemical potential automatically leads to an excess energy density, with the relation between $N_\nu$ and $\xi = \mu/T$ being
\begin{equation}
N_\nu = 3 + \sum_i \frac{30}{7} \left(\frac{\xi_i}{\pi}\right)^2+
\frac{15}{7} \left(\frac{\xi_i}{\pi}\right)^4,
\end{equation}
taking into account the possibility of different chemical potentials for the thee flavours.

Generically, large neutrino chemical potentials are excluded by the nucleosynthesis bound discussed previously. However, as was shown in \cite{Pastor:2008ti}, by tuning the initial asymmetries carefully it is possible to circumvent the nucleosynthesis bound and achieve a significant contribution to $N_\nu$ from non-zero chemical potentials.

Data from the Planck satellite which is currently observing the cosmic microwave background will have a much improved sensitivity to $N_\nu$, probably reaching $\sigma(N_\nu) \sim 0.2$ \cite{planck}. If the best-fit value remains at $N_\nu \sim 4$ Planck should therefore be able to provide evidence for new physics in this channel at fairly high significance.

\section{Future probes of neutrino mass in cosmology}

In the coming years a variety of different new experiments will probe cosmic structure with high precision across a wide range in
both scale and redshift.
Some probes are based on refinements of techniques already used. Examples of this are the Planck mission which is currently measuring the CMB
anisotropy, and upcoming galaxy surveys such as BOSS. Others will use completely new techniques, for example looking at structures at high redshift
using radio observations, or probe weak gravitational lensing of galaxies by cosmic large scale structure.

In this bewildering multitude of different experiments there are some reasonably simple guidelines which can be used to estimate the
sensitivity to neutrino parameters.
The underlying quantity measured by most of these experiments is the matter power spectrum. When probing for example inflationary models
higher order statistics such as the bispectrum or trispectrum can also be important, but for neutrino masses the power spectrum is normally
the relevant quantity.

The precision with which the power spectrum can be measured depends on a number of parameters. First of all, there is sample variance, i.e.\ the
fact that only a finite number of Fourier modes are contained within the survey volume. In some cases (like the CMB) this is a fundamental problem related to the horizon, but in most cases it is a technical problem because the effective survey volume is determined by the size of the telescope, the time available, and the luminosity of the objects studied. Beating down sample variance is best done using bright source objects to trace the underlying matter distribution. However, this leads to another fundamental problem, shot noise. The density of bright galaxies is low and therefore
small scale modes cannot be properly sampled. For galaxy surveys this is not a major concern because luminous red galaxies, such as used in the SDSS, are common enough that the power spectrum can be be measured to $k$ significantly larger than 0.1 $h$/Mpc.
For weak lensing surveys it is more of a concern because many more galaxies need to be measured to infer an average distortion in a given patch of the sky.

Finally, there can be serious systematic problems related to the extraction of the underlying matter power spectrum from a given measurement. In most cases the measurement is not of the matter density itself, but rather of some possibly biased tracer of the matter density. This is true for example in galaxy surveys where it is necessary to understand how the galaxy clustering power spectra for various types of galaxies are related to the underlying matter power spectrum. For Lyman-$\alpha$ forest measurements the problem is even worse because the measurement is of the column density of neutral hydrogen along lines of sight, a quantity which requires extensive modeling before a matter power spectrum estimate can be extracted. Similar problems will certainly face future measurements of temperature fluctuations in the 21-cm band.

The following subsections contain a brief review of the techniques which are currently used or under development. Note that all sensitivities are to the sum of neutrino masses, $\sum m_\nu$, and at 1$\sigma$ unless stated otherwise.

\subsection{The cosmic microwave background}

While the WMAP satellite has provided cosmic variance limited measurements of the temperature anisotropy up to $l \sim 500$, there are vast improvements to be made. The Planck mission \cite{planck} is currently measuring the CMB anisotropy and will provide cosmic variance limited temperature measurements up to $l \sim 2000$, as well as vastly better measurements of the polarisation anisotropy.
Even though the primary CMB signal is not highly sensitive to the neutrino mass, Planck will improve the WMAP constraint from 1.3 eV in the minimal
model to 0.4-0.5 eV \cite{Lesgourgues:2006nd,Perotto:2006rj,Hamann:2010pw}, and to 0.5-0.6 eV in more extended models. Even though this result is on par with the current constraint it is significantly more robust because it relies on only one data set, which is furthermore measured on scales large enough to make non-linear contributions subdominant.

However, even if such non-linear corrections are subdominant they are potentially important for measuring the neutrino mass.
Just as with any other source of radiation at high redshift, the CMB signal is affected by gravitational lensing along the line of sight.
For the CMB this effect can be diagnosed in a number of ways. For example it introduces non-Gaussian features in the otherwise Gaussian primary signal (see e.g.\ \cite{Lesgourgues:2005yv,Perotto:2006rj}). Another important effect is that it mixes $E$ and $B$ mode polarisation of the CMB \cite{Hu:2001tn,Hirata:2003ka}, and any experiment with sufficient sensitivity to the very weak $B$ mode signal should be able to identify this.

With lensing extraction Planck data alone should be able to reach a sensitivity of 0.1-0.15 eV \cite{Perotto:2006rj}, depending on the complexity of the model assumed. Further into the future, experiments highly sensitive to $B$-mode polarisation such as CMBPol \cite{Baumann:2008aq} might plausible reach a sensitivity of 0.04 eV.

\subsection{Galaxy surveys}

Galaxy surveys have up to now been the most direct way of measuring the matter power spectrum on intermediate and small scales, and therefore also the most direct probe of the suppression of fluctuation power caused by the presence of massive neutrinos. At present by far the largest spectroscopic survey is the SDSS, and, together with the WMAP CMB data, it provides an upper bound of approximately 0.6 eV on $\sum m_\nu$ \cite{Reid:2009xm}.

Galaxy redshift surveys measure the power spectrum of
 galaxy number density fluctuations, $P_g(k)$.  In turn, this power
 spectrum is related to the underlying  matter power spectrum $P(k)$
via
\begin{equation}
P_g(k)=b^2(k) P(k),
\end{equation}
where the bias parameter $b$ depends on both scale and on the type of galaxies surveyed. This has been shown to be a significant problem on some surveys and therefore emphasis has shifted towards basing surveys on luminous red cluster galaxies which constitute a fairly homogeneous sample. This is for example the case for the SDSS-LRG sample which was used to derive the current 0.6 eV upper bound.

A number of larger galaxy surveys will be carried out within the next decade and will increase sensitivity to neutrino mass significantly.
Given some survey design one can expect to measure the galaxy power spectrum
$P_g(k)$
up to a statistical uncertainty of
\cite{Tegmark:1997rp}
\begin{equation}
\label{eq:error}
\Delta P_g(k) = \sqrt{\frac{1}{2 \pi  \ w(k) \ \Delta \ln k}} \left[P_g(k) + \frac{1}{\overline{n}_g} \right].
\end{equation}
Here, $w(k) = (k/2 \pi)^3 \ V_{\rm eff}$, $V_{\rm eff}$
is the effective volume of the survey, $\overline{n}_g$ is the
galaxy number density, and $\Delta \ln k$ is the bin size at $k$ in $\ln k$-space. Future surveys will go deeper (and therefore have larger $\overline{n}_g$) and wider (and therefore increased $V_{\rm eff}$) leading to much smaller errors on the power spectrum.

The precision with which the power spectrum can, in principle, be measured is related to the survey volume because that is a measure of the number of independent Fourier modes available. On small scales precision is limited by shot noise, i.e.\ by the sparsity of galaxies.
However, in practice this is not the most significant problem on small scales. Rather, the usefulness of small scale data is limited by the fact that structures are non-linear. At $z = 0$ this effectively cuts away all data at $k > 0.1 \, h$/Mpc.

However, most of the upcoming surveys aim at measuring at higher redshift than the SDSS and therefore the problem of non-linearity will be somewhat alleviated.
In \cite{Hannestad:2007cp} a study of neutrino mass constraints was carried out for a number of proposed surveys combined with Planck data.
Very roughly, the HETDEX \cite{bib:hetdex}, WFMOS \cite{Glazebrook:2005ui}, or BOSS \cite{boss} surveys, together with Planck should push the sensitivity to about 0.2 eV at 95\% C.L., and a future space-based mission such as JDEM or EUCLID could yield a sensitivity of around 0.1 eV (95\% C.L.).

The major theoretical hurdles that need to be addressed in order to extract these sensitive limits are understanding the nonlinearities and bias. Simulations and cross-correlating with lensing surveys can help with these issues.

\subsection{Weak lensing surveys}

Weak gravitational lensing provides the only direct way to probe the cosmic matter distribution, as opposed to the distribution of some possibly biased tracer like galaxies. Perturbations in the matter density bend light rays from distant sources and leads to distortions in the shapes of objects such as galaxies. While galaxies are intrinsically elliptical the alignment of their axes should be close to random. Lensing on the other hand produced a coherent effect and by measuring the shape of many source galaxies and subsequently averaging it is possible to get an estimate of the weak lensing distortion field.

By measuring the weak lensing distortion (angular) power spectrum it is possible to probe the power spectrum of matter fluctuations between source and observer. Since it is at least in principle possible to measure the redshift of source galaxies it is furthermore possible to perform weak lensing tomography and probe the time evolution of the cosmic density field. It is impossible to obtain spectral redshifts for all source galaxies (the largest surveys will contain billions of such galaxies), but redshifts can be measured quite accurately by observing in different wavelength bands.

Although weak lensing surveys are still in their infancy they already provide some constraints on cosmology. At this stage the constraint is mainly just on the amplitude of matter fluctuations on intermediate scales. Data from the CFHT Legacy Survey has been used in combination with the WMAP 5-year data to provide an upper bound on the neutrino mass of $\sum m_\nu < 1.1$ at 95\% C.L. \cite{Ichiki:2008ye}.

Future weak lensing surveys from instruments such as the LSST \cite{:2009pq} will cover a large fraction of the sky and extend to redshifts of 1 or beyond. When combined with Planck data this will allow for very tight constraints on the neutrino mass. For example \cite{Hannestad:2006as} (see also \cite{kaplinghat}) has estimated that by binning the LSST data in 5 tomographic bins a sensitivity of 0.07 eV at 95\% C.L. can be reached. Proposed future surveys from space by experiments such as DUNE/Euclid may reach a comparable or better sensitivity \cite{Kitching:2008dp}.

One possibly severe problem with weak lensing surveys reaching this sensitivity is that, like CMB measurements, they measure the correlation function on a sphere, not in 3D. This means that they are sensitive to an integral along the line of sight of the underlying 3D fluctuation spectrum. This integration mixes small and large scale modes and therefore non-linear effects become important for weak lensing surveys even at relatively modest $l$-values. Since cosmological constraints can mainly be derived from moderate and large $l$-values it means that non-linear corrections must be under control at the percent level. For neutrinos this has already been achieved \cite{Brandbyge1,Brandbyge2,Brandbyge3}, but for example baryonic physics can contribute an uncertainty of 10\% or more on scale which are relevant for cosmological parameter extraction \cite{Jing:2005gm}.

\subsection{Lyman-$\alpha$ forest measurements}

An indirect way of measuring the matter power spectrum at high redshift consists of measuring absorption lines
from neutral hydrogen along the lines of sight to quasars. The width and amplitude of the Lyman-$\alpha$ absorption lines are related to the local hydrogen density and therefore also (albeit indirectly) related to the local
matter density.
An advantage of this method is that it can be used to probe high redshifts where structure are more linear. It can also probe much smaller scales than conventional galaxy surveys because it does not rely on a high density of bright objects such as LRGs.
Historically the method was first used to constrain neutrino masses in~\cite{Croft:2000hs}, and subsequently the large SDSS sample of absorption spectra was used in \cite{Seljak:2006bg} to infer an extremely stringent bound on $\sum m_\nu$ of 0.17 eV at 95\% C.L.

However, as has been discussed at length in \cite{Seljak:2006bg,Viel:2005ha,Viel:2010bn} the bound comes mainly from the fact that the small scale matter power spectrum normalisation is inferred in \cite{Seljak:2006bg} to be significantly higher than the one found from other cosmological data.
Using more up to date simulations which include neutrinos in the simulations \cite{Viel:2010bn} find a normalisation consistent with other probes and therefore a much less restrictive bound.

In the future there will be a number of new surveys such as BOSS \cite{boss} which will probe the Lyman-$\alpha$ forest in more detail. Together with more accurate modeling it seems likely that such surveys can reach a sensitivity in the 0.1 eV region when combined with CMB data.

\subsection{Cluster surveys}

Galaxy clusters, with masses around $10^{14}-10^{15} M_\odot$ are the largest gravitationally bound objects in the universe. Any change in the matter power spectrum on small scales shifts the mass of the largest object capable of forming before the onset of dark energy domination, and since neutrinos have a big impact on the power spectrum on these scales the presence of massive neutrinos has a noticeable impact on the halo mass function, i.e.\ the number density of halos with a given mass.
This point was discussed in detail in Section \ref{sec:hmf} (and for even more detail the reader is referred to \cite{brandbyge4}).

The biggest problem facing cluster surveys is that it is difficult to estimate the mass of a cluster based on simple observables such as their X-ray luminosity. Weak lensing measurements are more robust, but also significantly more time consuming. Fortunately, a number of large scale cluster surveys will become available in the not too distant future. For example the LSST survey \cite{lsst} will provide a sample of thousands of clusters. Hopefully this will also make it possible to make a much better calibration of such quantities as the mass-temperature relation.

If these issues can be reliably addressed, cluster surveys in combination with CMB data from Planck could potentially reach a sensitivity in the 0.04-0.07 eV range \cite{Wang:2005vr}.

\subsection{21-cm measurements}

Radio band observations of the 21-cm line of neutral hydrogen can potentially reach much higher redshifts than any
observations in the visible or infrared bands, and may provide access to information about the universe prior to reionization at $6 \lesssim z \lesssim 12$.
The brightness of a given patch of the universe in the 21-cm band depends on a number of factors and is given
approximately by \cite{fob}
\begin{equation}
\delta T_{21} = T_{21}-\bar{T}_{21} \sim 25 x_{\rm H} (1+\delta) \left(\frac{1+z}{10}\right)^{1/2} \left(1-\frac{T_\gamma}{T_s}\right)\left(\frac{H(z)/(1+z)}{dv_r/dr}\right),
\end{equation}
where $x_{\rm H}$ is the fraction of neutral hydrogen, $\delta$ is the density inhomogeneity, $T_s$ is the spin temperature, and $v_r$ is the radial velocity.
Thus, a measurement of $\delta T_{21}$ in principle provides information on $\delta$ and can be used to extract the matter power spectrum.

One huge advantage of the 21-cm radiation is that it covers an incredibly large volume, comparable to the entire Hubble volume. Furthermore, since it is a 3D measurement, as opposed to the CMB measurements which are on a sphere, there are many more independent Fourier modes than in any CMB measurements.

Given that the power spectrum can be reliably extracted from future observations, some of the upcoming or proposed radio experiments will be extremely sensitive to neutrino masses.
For example, the Square Kilometer Array (SKA) might constrain $\sum m_\nu$ to 0.02 eV, while the proposed Fast Fourier Transform Telescope (FFTT) \cite{tegmark2008} could achieve $\sigma(\sum m_\nu) \sim 0.0003$ eV \cite{mcquinn2006,mao2008}.

However, before this can be achieved there are some severe obstacles to be dealt with. The foreground contamination of highly redshiftet 21-cm radiation is enormous \cite{fob}. For example the galactic synchrotron emission is 4-5 orders of magnitude larger than the primary signal. There are additional problems with terrestrial contamination. It remains to be seen to which extent (if indeed any) 21-cm measurements can be used for precision cosmology.

\subsubsection{Summary of future probes}

As can be gauged from the above discussion there is cause for optimism on cosmological neutrino mass measurements. Many of the surveys discussed above will be realised within the next 5-7 years, and a number of different techniques with completely different systematics all have comparable potential sensitivities in the $\sigma(m_\nu) \sim 0.05$ eV range. This means that there will be many cross-checks available and that any possible detection of a neutrino mass will not rely on any single data set which could be dominated by unaccounted for systematics. Table~\ref{tab:future} shows a summary of the achievable sensitivities using various techniques.

Another question is what exactly will be learned about neutrino physics from these measurements. If neutrinos hierarchical or inverted hierarchical the sum of neutrino masses will be in the 0.05-0.1 eV range, and cosmological probes within the next decade will at most be able to provide a 1-2$\sigma$ evidence for non-zero mass. This also means that it is unlikely that cosmology will determine whether the normal or the inverted hierarchy is the correct model in that time frame. In order for such a determination, the formal sensitivity would have to be significantly better than 0.05 eV and so far 21-cm observations and very large scale lensing surveys seem the best path forward
(see \cite{Lesgourgues:2004ps,Jimenez:2010ev} for a more detailed discussion of probing the hierarchy with cosmology).
If neutrino masses are hierarchical or inverted hierarchical it is also unlikely that direct laboratory measurements will provide any significant evidence of neutrino mass within the next decade. Certainly beta decay experiments like KATRIN will remain far off the mark for the foreseeable future, whereas neutrinoless double beta decay experiments may or may not find evidence, depending on the nature of the hierarchy and exact Majorana phase structure.

However, it is also entirely possible that neutrino masses are degenerate in which case cosmology should provide a high significance detection within the next decade. Furthermore this possibility is interesting because both beta decay and neutrinoless double beta decay experiments hold the potential to achieve a positive detection. Having measurements of three different new neutrino mass observables, $m_{\nu_e}$, $m_{\beta \beta}$, and $\sum m_\nu$ will provide a plethora of new information on neutrino physics, including the values of Majorana phases.

\begin{table}[h]
\begin{center}
\begin{tabular}{|l|c|c|}
\hline
Probe &  Potential sensitivity (short term) & Potential sensitivity (long term)  \\
\hline
\hline
CMB & 0.4-0.6 & 0.4  \\
\hline
CMB with lensing & 0.1-0.15 & 0.04 \\
\hline
CMB + Galaxy Distribution & 0.2 & 0.05-0.1 \\
\hline
CMB + Lensing of Galaxies & 0.1 & 0.03-0.04 \\
\hline
CMB + Lyman-$\alpha$ & 0.1-0.2 & Unknown \\
\hline
CMB + Galaxy Clusters & - & 0.05 \\
\hline
CMB + 21 cm & - & 0.0003-0.1 \\
\hline
\end{tabular}
\caption{\label{tab:future}Future probes of neutrino mass, as well as their projected sensitivity to neutrino mass.
Sensitivity in the short term means achievable in approximately 5-7 years, while long term means 7-15 years.}
\end{center}
\vspace{-0.6cm}
\end{table}

\section{Detection of the cosmic neutrino background}

Standard model
physics likewise predicts the presence of a Cosmic Neutrino Background (C$\nu$B) with a well defined
temperature of $T_\nu \sim (4/11)^{1/3} T_\gamma$. While it remains undetected in direct experiments, the
presence of the C$\nu$B is strongly hinted at in CMB data. The homogeneous C$\nu$B component has been
detected at the 4-5$\sigma$ level in the WMAP data. Furthermore, this component is known to be
free-streaming, i.e. to have an anisotropic stress component consistent with what is expected from standard
model neutrinos (see
\cite{Bashinsky:2003tk,Trotta:2004ty,Bell:2005dr,DeBernardis:2008ys,Basboll:2008fx,Hannestad:2004qu,Friedland:2007vv}).
Finally the standard model neutrino decoupling history is also confirmed by Big Bang Nucleosynthesis (BBN),
the outcome of which depends on both the energy density and flavour composition of the C$\nu$B.

While this indirect evidence for the presence of a C$\nu$B is important, a direct detection remains an
intriguing, but almost impossible goal. The most credible proposed method is to look for a peak in beta
decay spectra related to neutrino absorption from the C$\nu$B \cite{Weinberg:1962zz,Cocco:2007za,Blennow:2008fh}, although
many other possibilities have been discussed
\cite{Weiler:1982qy,Stodolsky:1974aq,Gelmini:2004hg,Ringwald:2004np,Fodor:2002hy,Duda:2001hd,Langacker:1982ih,Cabibbo:1982bb}.

Near the electron endpoint energy there should be a small contamination from anti-neutrino capture from the background. This process is monoenergetic, and the electron energy should be $E_e \simeq Q + m_\nu$. Thus, the separation between the end of the beta spectrum and the absorption peak is $2 m_\nu$. Depending on the energy resolution and the source luminosity this may or may not be visible in future experiments. The problem was recently studied in more detail in \cite{Cocco:2007za}, with the conclusion that currently planned experiments like KATRIN are still several orders of magnitude away from a detection. Nevertheless, the method is in principle feasible and so far seems the most promising way to detect the neutrino background.

The neutrino absorption method was first investigated by Weinberg \cite{Weinberg:1962zz}, based on the
possibility that the primordial neutrino density could be orders of magnitude higher than normally assumed
due to the presence of a large chemical potential. Although a large chemical potential has been ruled out
because it is in conflict with BBN and CMB, the method may still work and
recently there has been renewed interest in detecting the C$\nu$B using beta unstable nuclei.

Although the direct detection of the C$\nu$B is already very challenging, one might speculate on the
possibility that in the more distant future anisotropies in the C$\nu$B will be detectable.

\subsection{Anisotropies in the cosmic neutrino background}

\begin{figure}
   \noindent
      \begin{center}
      \hspace*{-0.1cm}\includegraphics[width=0.8\linewidth]{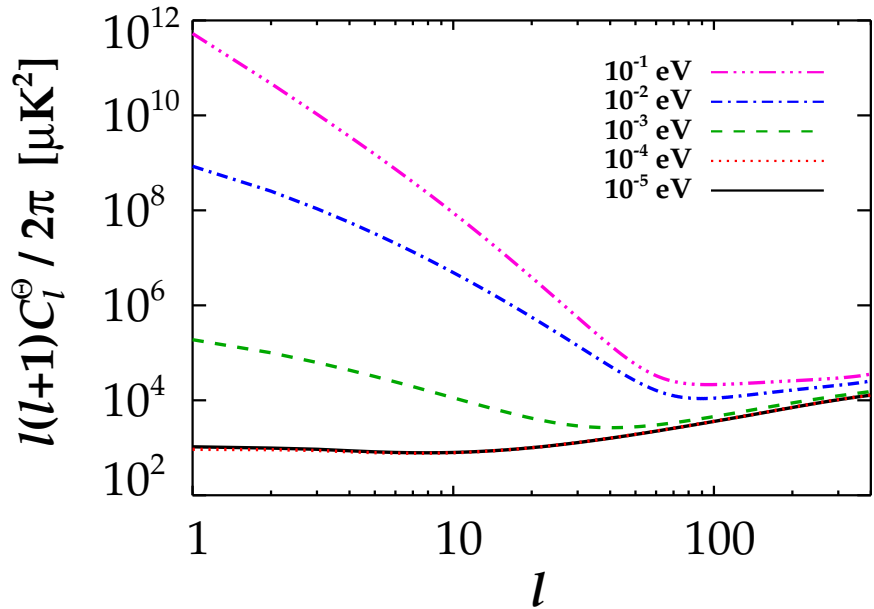}
      \end{center}
   \caption{Primary C$\nu$B spectrum for different neutrino masses.}
   \label{fig:transfer}
\end{figure}

In the more distant future it may become possible to not only detect the cosmic neutrino background, but also to measure the equivalent of the CMB anisotropy. Interestingly, for massive particles the cosmic background anisotropy becomes markedly different from the massless case. This was studied in \cite{hb10} for the case of light neutrinos. For neutrinos with average thermal velocity larger than $v \sim 1000$ km/s, very few neutrinos are bound, even in massive halos. Therefore, for neutrino masses below 0.1 eV a linear theory calculation of the cosmic neutrino background is appropriate. Furthermore it reveals some interesting features of how radiation anisotropies are generated. Since neutrinos are non-relativistic during a significant epoch in the late-time universe the gravitational source terms are much more important than for photons. Since these terms enter only through the lowest multipoles the effect is to very strongly enhance the low-$l$ part of the spectrum, an effect which can be seen in Fig.~\ref{fig:transfer}. If it ever becomes possible to measure the neutrino background, one would therefore mainly expect a low-$l$ signal associated with the most massive mass eigenstate.

\section{Discussion}

I have reviewed the topic of neutrino cosmology. The topic is quite diverse and I chose to put particular emphasis on how neutrinos affect structure formation in the universe while other interesting topics such as leptogenesis were left out.
At present neutrino physics presents perhaps the best example of how precision cosmology can be used to address particle questions which would normally be probed in laboratory experiments. The main example is the absolute mass of neutrinos: Absolute neutrino masses can only be measured at great difficulty in the lab, using either beta decay or neutrinoless double beta decay experiments. On the other hand, even small neutrino masses have a great impact on structure formation, and a sum of neutrino masses in the eV range would have produced a measurable suppression of small scale structure.

The current upper bound on the absolute neutrino mass is approximately $\sum m_\nu \lesssim 0.5$ eV, depending somewhat on the cosmological model and the data used. However, the bound cannot be pushed much beyond 1 eV even if several additional cosmological parameters are introduced.
In the future, a large number of new surveys will bring down the sensitivity to neutrino mass by an order of magnitude or more. For example weak gravitational lensing surveys will reach a sensitivity below 0.05 eV within the next decade.
Even this precision may not, however, be enough to guarantee a detection if neutrino masses are strongly hierarchical and follow the normal hierarchy.
To go below 0.05 eV at high significance will likely require future 21-cm surveys or possibly a new generation of space based weak lensing measurements. The timescale for such measurements is likely to be 15-20 years or more.
However, the theoretical lower limit to cosmological neutrino mass measurements is extremely low, and certainly low enough for a measurement of the individual neutrino masses even in the normal hierarchy.
At this stage it seems likely that the first positive detection of a non-zero neutrino mass (other than from oscillation experiments) will come from cosmology, and in any case cosmology will remain a powerful laboratory for neutrino physics which is complementary to laboratory experiments.

\section*{Acknowledgements}

Use of computing resources from the Danish Center for Scientific
Computing (DCSC) is acknowledged. I wish to thank the INFN Padova for hospitality during the completion of this work.


\section*{References} 

\end{document}